\documentclass[a4paper, superscriptaddress, amsfonts, amssymb, amsmath, reprint, showkeys, nofootinbib, twoside,aps,prl]{revtex4-1}
\usepackage[english]{babel}
\usepackage[utf8]{inputenc}
\usepackage{comment}
\usepackage{xcolor}
\usepackage{bm}
\usepackage{graphicx}
\usepackage{float}
\usepackage{amsmath}
\usepackage{amsfonts,amssymb,amsthm}
\usepackage[colorinlistoftodos, color=green!40, prependcaption]{todonotes}
\usepackage{amsthm}
\usepackage{mathtools}
\usepackage{physics}
\usepackage{xcolor}
\usepackage{graphicx}
\usepackage[left=23mm,right=13mm,top=35mm,columnsep=15pt]{geometry} 
\usepackage{adjustbox}
\usepackage{placeins}
\usepackage[T1]{fontenc}
\usepackage{lipsum}
\usepackage{csquotes}
\usepackage[pdftex, pdftitle={Article}, pdfauthor={Author}]{hyperref} 

\usepackage{geometry}
\begin{document}
\title{Decomposing Multivariate Information Rates in Networks of Random Processes}

\author{Laura Sparacino}
    \affiliation{Department of Engineering, University of Palermo, Palermo, Italy}
\author{Gorana Mijatovic}
    \affiliation{Faculty of Technical Sciences, University of Novi Sad, Serbia}
\author{Yuri Antonacci}
    \affiliation{Department of Engineering, University of Palermo, Palermo, Italy}
\author{Leonardo Ricci}
    \affiliation{Department of Physics, University of Trento, Italy}
\author{Daniele Marinazzo}
    \affiliation{University of Ghent, Belgium}
\author{Sebastiano Stramaglia}
    \affiliation{University of Bari Aldo Moro and Istituto Nazionale
di Fisica Nucleare, Sezione di Bari, Italy}
\author{Luca Faes}
    \email[Correspondence email address: ]{luca.faes@unipa.it}
    \affiliation{Department of Engineering, University of Palermo, Palermo, Italy}
    \affiliation{Faculty of Technical Sciences, University of Novi Sad, Serbia}

\begin{abstract}
The Partial Information Decomposition (PID) framework has emerged as a powerful tool for analyzing high-order interdependencies in complex network systems. However, its application to dynamic processes remains challenging due to the implicit assumption of memorylessness, which often falls in real-world scenarios. In this work, we introduce the  framework of Partial Information Rate Decomposition (PIRD) that extends PID to random processes with temporal correlations.
By leveraging mutual information rate (MIR) instead of mutual information (MI), our approach decomposes the dynamic information shared by multivariate random processes into unique, redundant, and synergistic contributions obtained aggregating information rate atoms in a principled manner. To \textcolor{black}{concretely implement this idea}, we define a pointwise redundancy rate function based on the minimum MI principle applied locally in the frequency-domain representation of the processes. 
The framework is validated in benchmark simulations of Gaussian systems, demonstrating its advantages over traditional PID in capturing temporal correlations and showing how the spectral representation may reveal scale-specific higher-order \textcolor{black}{interactions} that are obscured in the time domain.
Furthermore, we apply PIRD to a physiological network comprising cerebrovascular and cardiovascular variables, revealing frequency-dependent redundant information exchange during a protocol of postural stress.
Our results highlight the necessity of accounting for the full temporal statistical structure and spectral content of vector random processes to meaningfully perform information decomposition in network systems with dynamic behavior such as those typically encountered in neuroscience and physiology.

\end{abstract}


\maketitle

\section{\label{sect:Intro}Introduction}
In the growing research area of Network Science \cite{barabasi2013network}, identifying and untangling the multifaceted many-body interactions occurring in complex network systems composed by several interconnected units has become a crucial task. Different approaches have been developed to deepen our understanding of interactions in complex networks, and have been exploited in applicative fields ranging from neuroscience and biology to sociology and engineering. Advances have come from the utilization of many disparate tools such as graph theory and network analysis \cite{newman2003structure}, dynamical systems approaches including nonlinear methods, chaos theory, synchronization and coupled oscillator models \cite{strogatz2015nonlinear,boccaletti2002synchronization,pikovsky2003synchronization,ricci2021experimental}, tools for the analysis of higher-order networks, hypergraph models and simplicial complexes \cite{courtney2016generalized,battiston2021physics,santoro2023higher}, as well as methods rooted in the application of information theory to network models and time series data \cite{mcgill1954interaction,sole2004information,timme2014synergy,novelli2021inferring,perinelli2018correlation,perinelli2021relationship}.

In this context, partial information decomposition (PID) has been developed as a comprehensive framework designed to understand how information is distributed in multivariate systems: the foundational work by Williams and Beer \cite{williams2010nonnegative} introduced PID as a method to decompose multivariate information in non-negative terms, addressing limitations of traditional measures like the interaction information \cite{mcgill1954interaction} which can yield both positive and negative values often obscuring the interpretation of informational relationships. Considering a \textit{target} random variable and a set of \textit{source} variables, the mathematical redundancy lattice structure defined for the PID \cite{williams2010nonnegative} identifies a set of atoms whose associated partial information (PI) amounts constitute the building blocks of the analyzed multivariate information shared between the target and the sources, quantified by the mutual information (MI). 
Moreover, to overcome the limitation that the number of atoms grows super-exponentially with the number of source variables \cite{gutknecht2021bits}, refinements have been introduced whereby the PI atoms are aggregated meaningfully to highlight how the MI is distributed among the sources \cite{rosas2020reconciling}; these refined approaches provide a coarse-grained decomposition with a small number of atoms that scale gracefully with the system size, highlighting the \textit{unique} information exclusively available from each source, the \textit{redundant} information obtained from at least two different sources, and the \textit{synergistic} information revealed only when multiple sources are considered simultaneously.

Thanks to the peculiarities described above, the PID framework has become widely adopted as a main tool to assess high-order interdependencies among the units of network datasets collected in several applicative fields of physics, engineering and life sciences where understanding the specific contributions of individual variables to the collective information is of fundamental importance \cite{wibral2017partial, cang2020inferring, rosas2020reconciling, luppi2022synergistic, wollstadt2023rigorous, dissanayake2024quantifying}. For instance, PID has been used to analyze neural information processing dynamics, revealing how different brain regions encode stimuli both redundantly and synergistically \cite{kay2022comparison}. In machine learning, PID has been exploited in feature selection to distinguish between information that is redundant across features and information that is uniquely informative \cite{barnett2009partial, taylor2018partial,wollstadt2023rigorous,bara2024partial}. Recent applications have highlighted its value for network analysis also in computational neuroscience, biology and physiology \cite{luppi2024information,sherrill2021partial,faes2021information}.

However, in spite of the universality of the problem posed by PID, the underlying analytical framework presents some inherent limitations that restrict its unambiguous utilization in different contexts. A first issue was recognized since the inception of PID \cite{williams2010nonnegative} noting that the information atoms of
unique, redundant, and synergistic information cannot be
defined using classical information theory, but rather require the introduction of new axioms whose definition is not yet universally accepted.
Consequently, several axiomatic definitions of redundancy have been proposed so far that differ depending on the philosophy followed to satisfy the desired properties (e.g., decision- \cite{pakman2021estimating}, game- \cite{ince2017measuring}, information-theoretic \cite{makkeh2021introducing}), on the nature (continuous \cite{barrett2015exploration, pakman2021estimating, ehrlich2024partial} or discrete \cite{williams2010nonnegative, bertschinger2014quantifying, ince2017measuring}) of the analyzed variables, and on assumptions made about their distribution (e.g., Gaussian \cite{barrett2015exploration}). Popular and simple approaches implement the so-called minimum MI (MMI) PID schemes, whereby redundancy is defined for a given atom as the minimum of the MI (or the specific MI) shared between the target (or a specific state of it) and each source \cite{williams2010nonnegative,barrett2015exploration}. However, these schemes are limited in the fact that they quantify the minimum amount of information that all variables carry but do not require that such information is the same for all variables \cite{harder2013bivariate,bertschinger2014quantifying,griffith2015quantifying,ince2017measuring}. To overcome this limitation, less conservative approaches have been proposed which typically define redundancy at the local or pointwise level, rather than at the level of ensemble averages \cite{ince2017measuring,makkeh2021introducing,ehrlich2024partial}; these approaches allow defining more refined redundancy quantities, but suffer in their turn from the limitation of potentially yielding negative information atoms (essentially because local MI can be negative), which hinders a strightforward interpretation of the results. Thus, the definition of a proper operalization of PID, merging computability and interpretability, is still an open problem in information theory.

Another crucial issue, which is the main question addressed in this paper, is how to apply PID to random processes with temporal statistical structure.
In fact, although specifically defined for random variables, PID is often needed in practice to analyze multivariate time series whose most proper statistical representation is the vector random process. Then, since both the target random process and the set of source processes constitute collections of 
random variables, utilization of the PID in such dynamic case is not straightforward as it implies an arbitrary selection of the variables to be extracted from each process.
In the literature, PID has been applied to the variables sampling the processes at the same time, performing a so-called "static PID", under the implicit assumption that the processes are stationary and memoryless (i.e., composed by independent and identically distributed - i.i.d. - variables).
However, the i.i.d. assumption is typically not tested in practice and is often violated in applications of information decomposition where the analyzed data exhibit nontrivial temporal correlations \cite{kay2022comparison, varley2023information, varley2023multivariate}.
Alternatively, the PID has been applied to random processes by selecting the variables to use in order to decompose the joint information transferred from all sources to the target \cite{krohova2019multiscale,luppi2022synergistic,koccillari2023behavioural,luppi2024synergistic}. Specifically, taking the present of one predefined process as the target variable and the past histories of all other processes as (vector) source variables, and conditioning the MI between target and sources on the past of the predefined process, yields the PID of the popular transfer entropy (TE) measure of information transfer \cite{schreiber2000measuring}.
Nevertheless, although it considers the temporal statistical structure of the multivariate process, the PID of the TE cannot account for instantaneous interactions among the processes, nor for causal interactions occurring in the causal direction from the target to the sources. Hence, the current applications of PID to random processes provide only a partial, often misleading view of the interactions among the units of dynamic network systems. 

\textcolor{black}{To deal with the significant issues outlined above}, the present work introduces a framework for the decomposition of the information shared dynamically between the target and the source units composing the analyzed network of random processes. Our idea is to shift the paradigm of PID from the use of random variables to the use of random processes as building blocks of information decomposition: leveraging information rate quantities in place of standard information quantities, we replace the MI between random variables with the MI rate (MIR) between random processes, and use the same lattice backbone of PID to implement the so-called \textit{partial information rate decomposition} (PIRD).
PIRD dissects the information shared per unit of time between the designed target random process and the set of source processes, quantified by the MIR, into PI rate atoms adopting a full PID perspective, or into unique, redundant and synergistic information rates adopting a coarse-graining perspective.
\textcolor{black}{The PIRD is then completed by} introducing the new information-theoretic measure of \textit{redundancy rate} which generalizes the MIR over the lattice. Among several possible definitions, the redundancy rate is here formulated \textcolor{black}{for the first time} following a pointwise approach implemented in the frequency domain, i.e. quantifying the concept of redundancy among iso-frequency oscillatory components of the analyzed processes. \textcolor{black}{Specifically, the redundancy rate is defined first at each specific frequency applying the minimum MI principle \cite{barrett2015exploration} to the spectral MIR between the target and source processes, and then in the time domain by integrating over all frequencies. This allows to retrieve a non-negative decomposition of the MIR, both locally in frequency and globally in the time domain.}
\textcolor{black}{Furthermore, a computationally reliable implementation of the framework is provided for the case of linear Gaussian random processes for which the spectral MIR can be defined from the power spectral density matrix \cite{geweke1982measurement, chicharro2011spectral}.}
The feasibility of \textcolor{black}{this implementation is tested} exhaustively in benchmark simulations of linearly interacting processes, where we illustrate the benefits of PIRD over the static PID in the presence of non-trivial temporal correlations, investigate the differences with the PID applied to the TE, and highlight the peculiarity of the spectral approach which allows to perform PID restricting the analysis to predetermined frequency bands with practical meaning. The latter property is then exploited in the applicative context of Network Physiology \cite{ivanov2021new} reporting the application of the coarse-grained PIRD to a physiological network comprising cerebrovascular, cardiovascular and respiratory time series analyzed in patients prone to develop postural-related syncope.

\section{\label{sect:PID_variables}Decomposition of multivariate information}
Let us consider a static network system composed of $M+1$ nodes $\mathcal{V} = \{ \mathcal{T},\mathcal{S}_1,\ldots,\mathcal{S}_M \}$, where the activity at each node is described in terms of random variables $T,S_1,\ldots,S_M$, with $T$ assumed as the \textit{target} variable and $\textcolor{black}{\textbf{S}} = \{ S_1,\ldots,S_M \}$ as the vector of \textit{sources}. 
The identification of a target and several source variables naturally leads to use directed approaches to assess and decompose the multivariate information shared in the system, i.e., the mutual information (MI) between $T$ and \textcolor{black}{$\textbf{S}$} defined as
\begin{equation}
   I(T;\textcolor{black}{\textbf{S}})=\mathbb{E} \left[ \log \frac{p(t,\textcolor{black}{\textbf{s}})}{p(t)p(\textcolor{black}{\textbf{s}})} \right],
   \label{MutualInformation}
\end{equation}
where $p(\cdot,\cdot)$ and $p(\cdot)$ denote joint and marginal probability, with $\mathbb{E}[\cdot]$ being the expectation operator.
Directed approaches to information decomposition expand the MI in (\ref{MutualInformation}) into terms related to the contributions that the individual sources $S_i$, $i=1,\ldots,M$, or a collection thereof, share with the target $T$. Such contributions are related in non-trivial ways to the marginal MI between each source and the target, $I(T;S_i)$. In this context, the partial information decomposition (PID) introduced by Williams and Beer \cite{williams2010nonnegative} provides one of the most popular approaches to study multivariate systems through the lens of a directed decomposition of multivariate information.

\subsection{\label{subsect:PID_variables_PID}Partial Information Decomposition and the redundancy lattice}
\begin{figure*}
\centering
\includegraphics[scale=1.1]{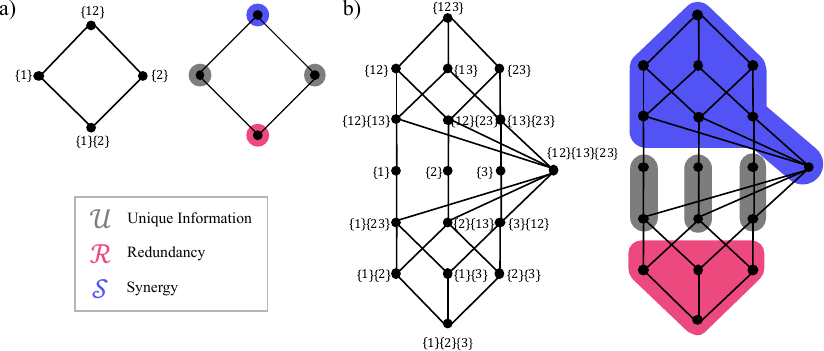}
\caption{\textbf{Standard and coarse-grained partial information decomposition, superimposed on the redundancy lattices for $M$ sources.}
\textbf{(a)} Standard PID on the redundancy lattice for 3 variables ($M=2$). The alphabet of source combinations is $\mathcal{A}= \{ \{1\}\{2\}, \{1\}, \{2\}, \{12\} \}$, where $\{1\}$ denotes $S_1$ and $\{2\}$ denotes $S_2$. The MI between the target and the set of sources is decomposed into a redundant (magenta), a synergistic (blue) and two unique (grey) contributions, which are exclusively provided by different atoms of the lattice. 
\textbf{(b)} Standard and coarse-grained $1^{st}$ order PID on the redundancy lattice for 4 variables ($M=3$). The alphabet of source combinations is $\mathcal{A}= \{ \{1\}\{2\}\{3\}, \{1\}\{2\}, \{1\}\{3\}, \{2\}\{3\}, \ldots, \{123\} \}$, where $\{i\}$ denotes $S_i$, $i=1,\ldots,M$. The MI between the target and the set of sources is decomposed into a redundant (magenta), a synergistic (blue) and three unique (grey) contributions, which are exclusively provided by different set of atoms of the lattice.}
\label{fig:fig_lattice}
\end{figure*}

Considering the target variable $T$ and the vector source variable \textcolor{black}{$\textbf{S}$}, the mathematical \textit{redundancy lattice structure} defined for the PID \cite{williams2010nonnegative} identifies a set of atoms whose associated partial information (PI) amounts constitute the building blocks of the analyzed multivariate information $I(T;\textcolor{black}{\textbf{S}})$.
The lattice is identified by the collection $\mathcal{A}$ of all subsets of sources such that no source is a superset of any other, i.e., the set of anti-chains formed from the indices of the sources in \textcolor{black}{$\textbf{S}$} under set inclusion \cite{williams2010nonnegative} 
(see Fig. \ref{fig:fig_lattice}a,b (left) for the cases of $M=2$ and $M=3$ sources; e.g, $\mathcal{A}=\{\{1\}\{2\},\{1\},\{2\},\{12\}\}$ if $M=2$). Formally, the PID expands (\ref{MutualInformation}) as
\begin{equation}
    I(T;\textcolor{black}{\textbf{S}}) = \sum_{\alpha \in \mathcal{A}} I^{\delta}(T;\textcolor{black}{\textbf{S}}_{\alpha}),
    \label{PID_atoms}
\end{equation}
where $I^{\delta}(\cdot;\cdot)$ is the PI function defined over the atoms of the lattice, $\alpha=\{\alpha_1,\ldots,\alpha_J\} \in \mathcal{A}$, and $\textcolor{black}{\textbf{S}}_{\alpha}$ denotes the set of subsets of source variables indexed by the $\alpha^{\mathrm{th}}$ atom, with $\alpha_j\subseteq \{1\cdots M\}$, $\textcolor{black}{\textbf{S}}_{\alpha_j} \subseteq \textcolor{black}{\textbf{S}}$, $j=1,\ldots,J$.

To complete the PID besides the basic statement in (\ref{PID_atoms}) it is necessary to provide a set of so-called consistency equations which, relating atoms to mutual information, allow to derive the PI terms \cite{gutknecht2021bits}. The main consistency equations state that the marginal MI terms involving any individual source variable $S_i$ are constructed additively by summing the information of the atoms positioned at the level $\{i\}$ and downwards in the lattice, i.e.,
\begin{equation}
    I(T;S_i)=\sum_{\beta \preceq \{i\}} I^{\delta}(T;\textcolor{black}{\textbf{S}}_{\beta}),
    \label{marginalMI}
\end{equation}
where $\preceq$ identifies precedence based on the partial ordering imposed by the lattice structure \cite{williams2010nonnegative}. 
Moreover, as the equations (\ref{PID_atoms}) and (\ref{marginalMI}) do not suffice to solve the PID problem because they provide a number of constraints lower than the number of information atoms to be computed (i.e., $M+1<|\mathcal{A}|$, where $|\cdot|$ indicates cardinality), to complete the PID it is necessary to define a so-called \textit{redundancy function}, here denoted as $I^{\cap}(\cdot;\cdot)$, which generalizes the MI over the lattice. The redundancy function extends (\ref{marginalMI}) to each atom $\alpha \in \mathcal{A}$, fulfilling 
\begin{equation}
    I^{\cap}(T;\textcolor{black}{\textbf{S}}_{\alpha})=\sum_{\beta \preceq \alpha} I^{\delta}(T;\textcolor{black}{\textbf{S}}_{\beta}),
    \label{redundancy_function}
\end{equation}
where $\beta$ represents the atoms preceding or equal to $\alpha$ and structurally connected to it in the lattice.
Finally, the information associated to all atoms can be retrieved, starting from the knowledge of the redundancy function, either iteratively as
\begin{equation}
    I^{\delta}(T;\textcolor{black}{\textbf{S}}_{\alpha})=I^{\cap}(T;\textcolor{black}{\textbf{S}}_{\alpha})-\sum_{\beta \prec \alpha} I^{\delta}(T;\textcolor{black}{\textbf{S}}_{\beta}),
    \label{delta_from_red}
\end{equation}
or in a compact way via  M\"{o}bius inversion of (\ref{redundancy_function}) \cite{williams2010nonnegative}. Importantly, while the redundancy value of an atom $\alpha$, $I^{\cap}(T;\textcolor{black}{\textbf{S}}_{\alpha})$, measures the total amount of redundant information shared by all the sources included in that atom, the PI measures the unique information contributed only by that atom.

At this point is worth noting that, although the PID is solved once a redundancy function is assigned, there is no consensus in the literature about how to define such function unequivocally. This is an active area of research, with several axiomatic definitions of redundancy proposed so far that differ depending on the philosophy followed to satisfy the desired properties (e.g., decision- \cite{pakman2021estimating}, game- \cite{ince2017measuring}, information-theoretic \cite{makkeh2021introducing}), on the nature (continuous \cite{barrett2015exploration, pakman2021estimating, ehrlich2024partial} or discrete \cite{williams2010nonnegative, bertschinger2014quantifying, ince2017measuring}) of the analyzed variables, and on assumptions made about their distribution (e.g., Gaussian \cite{barrett2015exploration}). A popular and simple approach is the so-called minimum MI (MMI) PID, whereby redundancy is defined for the atom $\alpha=\{\alpha_1,\ldots,\alpha_J\}$ as the minimum of the information shared between the target and each individual information component \cite{barrett2015exploration}:
\begin{equation}
    I^{\cap}(T;\textcolor{black}{\textbf{S}}_{\alpha}) \textcolor{black}{= I^{\mathrm{MMI}}(T;\textbf{S}_{\alpha}) := }\min\limits_{j=1,\ldots,J} I(T;\textcolor{black}{\textbf{S}}_{\alpha_j});
    \label{redundancy_MMI}
\end{equation}
for instance, considering $M=3$ sources and the atom $\alpha=\{\{1\},\{23\}\}$, the redundancy becomes the minimum between $I(T;S_1)$ and $I(T;S_2,S_3)$. 
\textcolor{black}{Although it tends to overestimate the amount of redundancy expected by intuition \cite{timme2014synergy, griffith2015quantifying,harder2013bivariate}}, the MMI PID is widely used thanks to its simplicity and generality; in particular, it is generally applied in case of Gaussian data, for which it subsumes several of the previously proposed PID schemes \cite{barrett2015exploration}. We will adopt \textcolor{black}{a modified version of} this approach in the practical implementation of the redundancy rate for random processes defined in Sect. "Decomposition of Multivariate Information".

\subsection{Coarse-grained PID}

An important aspect with practical relevance is that, as an alternative to (\ref{PID_atoms}), the PID can be formulated in a meaningful way by making explicit the \textit{unique information} that each source $S_m$ holds about the target $T$ ($m=1,\ldots,M$), the \textit{redundant information} that all source variables in \textcolor{black}{$\textbf{S}$} hold about $T$, and the \textit{synergistic information} about $T$ that only arises from knowing all the sources $S_1,\ldots,S_M$. This corresponds to expand multivariate information as
\begin{equation}
    I(T;\textcolor{black}{\textbf{S}}) =  \sum_{m=1}^{M} \mathcal{U}(T;S_m) + \mathcal{R}(T;\textcolor{black}{\textbf{S}}) + \mathcal{S}(T;\textcolor{black}{\textbf{S}}),
    \label{PID_contributions}
\end{equation}
where
\begin{equation}
    \mathcal{U}(T;S_m) = I(T;S_m) - \mathcal{R}(T;\textcolor{black}{\textbf{S}}).
    \label{UI_MI_equivalence}
\end{equation}
This approach provides a \textit{coarse-grained} decomposition with a small number of atoms that scale gracefully with the system size: while the full PID (\ref{PID_atoms}) yields a number of atoms $|\mathcal{A}|$ that grows super-exponentially with $M$ like the Dedekind numbers \cite{gutknecht2021bits}, the coarse-grained PID (\ref{PID_contributions}) decomposes the multivariate information into exactly $M+2$ quantities.
In particular, the two formulations coincide when $M=2$ source variables are considered, yielding $\mathcal{R}(T;\textcolor{black}{\textbf{S}})=I^{\delta}(T;\textcolor{black}{\textbf{S}}_{\{1\}\{2\}})$, $\mathcal{U}(T;S_1)=I^{\delta}(T;S_{\{1\}})$, $\mathcal{U}(T;S_2)=I^{\delta}(T;S_{\{2\}})$, and $\mathcal{S}(T;\textcolor{black}{\textbf{S}})=I^{\delta}(T;\textcolor{black}{\textbf{S}}_{\{12\}})$ (Fig. \ref{fig:fig_lattice}a, right). On the other hand, when $M \geq 3$ the coarse-graining is implemented by summing the PI of some of the atoms in (\ref{PID_atoms}).
This issue has been addressed in \cite{rosas2020reconciling}, where the construction for $M=2$ was generalized to $M=3$ sources through the so-called $k^{\mathrm{th}}$ order \textit{ coarse-grained} PID, which preserves the intuitive meaning that synergy, redundancy, and unique information have for $M=2$ sources (Fig. \ref{fig:fig_lattice}b, right).
Specifically, considering $k=1$, the $1^{\mathrm{st}}$-order synergy, $\mathcal{S}(T;\textcolor{black}{\textbf{S}})$, corresponds to the information about the target that is provided by the whole \textcolor{black}{$\textbf{S}$} but is not contained in any subset of sources when considered separately from the rest (atoms surrounded by the blue shade in Fig. \ref{fig:fig_lattice}).
Similarly, the $1^{\mathrm{st}}$-order redundancy, $\mathcal{R}(T;\textcolor{black}{\textbf{S}})$, is the information held by at least two different groups of size $1$ (magenta in Fig. \ref{fig:fig_lattice}).
Finally, the $1^{\mathrm{st}}$-order unique information provided by the $m^{\mathrm{th}}$ source, $\mathcal{U}(T;S_m)$, $m=1,\ldots,M$, is the information that $S_m$ has access to and no other subset of parts has access to on its own, although bigger groups of other parts may have (grey in Fig. \ref{fig:fig_lattice}). 

\subsection{\label{subsect:PID_variables_linear_formulation}Linear parametric formulation}
In the case in which the observed variables have a joint Gaussian distribution, the PID can be performed by exploiting linear parametric regression models. Specifically,  if $T \sim \mathcal{N}(m_T,\sigma^2_{T})$ and $S_i \sim \mathcal{N}(m_{S_i},\sigma^2_{S_i})$, $i=1,\ldots,M$, the target $T$ and the source $S_i$ are related by the following linear regression model:
\begin{equation}
    T = a S_i + b + U_i,
    \label{VAR_model_static}
\end{equation}
where $T$ is predicted using the coefficient $a$ weighing the regressor $S_i$, $b$ is the constant term and $U \sim \mathcal{N}(0,\sigma^2_{U_i})$ is a zero-mean Gaussian innovation with variance $\sigma^2_{U_i}$, uncorrelated with $S_i$.
Then, the MI between $T$ and $S_i$ can be estimated exploiting the relation between entropy and variance valid for Gaussian variables \cite{barrett2010}. Specifically, expressing the entropy of the predicted variable $T$ as $H(T) = \frac{1}{2} \log{( 2 \pi e \sigma^2_{T})}$, and the conditional entropy of the predicted variable $T$ given the predictor $S_i$ as $H(T|S_i) = \frac{1}{2} \log{( 2 \pi e \sigma^2_{U_i} )}$, yields to compute the MI as
\begin{equation}
    I(T;S_i) = H(T)-H(T|S_i) = \frac{1}{2} \log{\biggr( \frac{\sigma^2_{T}}{\sigma^2_{U}} \biggr)}.
    \label{MI_linear}
\end{equation}
Eq. (\ref{MI_linear}) can be applied to any combination of sources, allowing then to compute the redundancy function using (\ref{redundancy_MMI}) and the PI of each atom using (\ref{delta_from_red}), from which the coarse-grained terms are derived using (\ref{PID_contributions}).

\section{\label{sectPIRDprocesses}Decomposition of multivariate information rates} 
\begin{figure}
    \centering
    \includegraphics
    {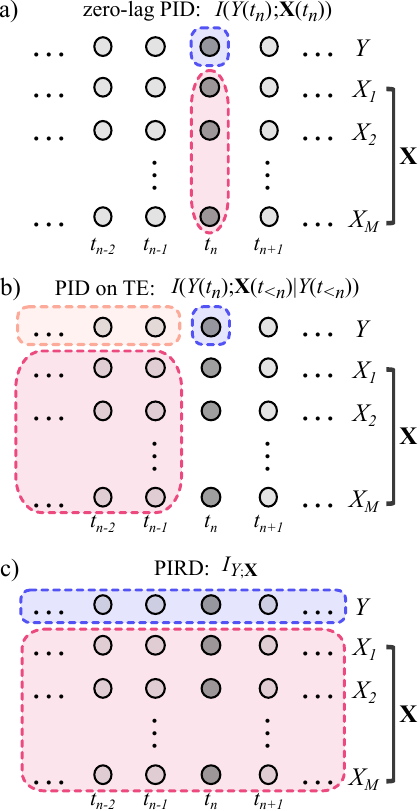}
    \caption{
    \textcolor{black}{\textbf{Selection of the variables to be used in PID schemes applied to a network of random processes.}
    Each panel depicts the random variables (circles) sampling the vector random process $\{Y,\textbf{X} \} = \{Y, X_1,\ldots,X_M \}$ at specific time points.
    \textbf{a)} The \textit{zero-lag PID} is applied to the MI between variables sampling the processes at the same time $t_n$, taking $Y(t_n)$ (blue) and $\textbf{X}(t_n)$ (magenta) as target and source variables.
    \textbf{b)} The PID applied to the \textit{joint TE} considers the variable sampling the process $Y$ at the current time $t_n$, $Y(t_n)$,  as target variable (blue) and the variables sampling the past of the process $\textbf{X}$ at times preceding $t_n$, $\textbf{X}(t_{<n})$, as the set of sources (magenta), and decomposes a conditional MI where the past of $Y$, $Y(t_{<n})$, is used for conditioning (pink).
    \textbf{c)} The \textit{PIRD} considers the whole processes as building blocks for the decomposition, and is applied to the MIR between the target process $Y$ (blue) and the source processes in $\textbf{X}$ (magenta).}
    }
    \label{fig:intro}
\end{figure}

Let us consider a dynamic network system composed of $M+1$ nodes, $\mathcal{Z}=\{ \mathcal{Y},\mathcal{X}_1,\ldots,\mathcal{X}_M \}$, where the activity at each node is described in terms of the vector random process $\textcolor{black}{\textbf{Z}}=\{Y,X_1,\ldots,X_M\}$, with $Y$ assumed as \textit{target} and $\textcolor{black}{\textbf{X}} = \{ X_1,\ldots,X_M \}$ as the vector of \textit{sources}. 
Here we consider discrete-time random processes intended as time-ordered vector random variables: e.g., \textcolor{black}{$Y(t_n)$} is the variable sampling the target process at the time $t_n$, where $n \in \mathbb{Z}$ is the time index; typically, $t_n=n\Delta t$, where $\Delta t$ is the sampling period ($f_s=\Delta t^{-1}$ is the sampling frequency)\textcolor{black}{; see Fig. \ref{fig:intro} for a graphical representation}.

Since each random process is a collection of random variables, the application of PID in the dynamic case is not straightforward as it implies an arbitrary selection of the variables to be extracted from each process. The most intuitive choice is to apply the PID to the variables sampling the processes at the same time $t_n$, setting \textcolor{black}{$T=Y(t_n)$} and \textcolor{black}{$\textbf{S}=\textbf{X}(t_n)$} as target and source variables and thus decomposing the static MI \textcolor{black}{$I(Y(t_n);\textbf{X}(t_n))$} through a redundancy function based on zero-lag MI terms (e.g., $I^{\cap}(T;S_i)=\textcolor{black}{I(Y(t_n);X_{i}(t_n))}$) according to the formalism presented in Sect. "Partial Information Decomposition and the redundancy lattice" \textcolor{black}{(Fig. \ref{fig:intro}a)}.
An alternative approach is to decompose the joint transfer entropy (TE) from all sources to the target, which is defined as \textcolor{black}{$T_{\textbf{X}\rightarrow Y}=I(Y(t_n);\textbf{X}(t_{<n})|Y(t_{<n}))$ \cite{schreiber2000measuring}, where $\textbf{X}(t_{<n})=\{\textbf{X}(t_{n-1}),\textbf{X}(t_{n-2}),\ldots\}$ and $Y(t_{<n})=\{Y(t_{n-1}),Y(t_{n-2}),\ldots\}$} denote the (potentially infinite-dimensional) vectors representing the past history of the source and target processes; in this case the PID is applied setting \textcolor{black}{$T=Y(t_n)$} and \textcolor{black}{$\textbf{S}=\textbf{X}(t_{<n})$} as target and source variables, and the redundancy function is a conditional MI (e.g., $I^{\cap}(T;S_i)=\textcolor{black}{I(Y(t_n);X_i(t_{<n})|Y(t_{<n}))}$\textcolor{black}{, Fig. \ref{fig:intro}b)}.

Although these two alternative applications of PID to random processes are common \cite{kay2022comparison,varley2023information,koccillari2023behavioural,luppi2024information}, they provide only a partial view of the dynamic interactions among the processes and are based on implicit assumptions which are often not fulfilled by dynamic network systems.
The zero-lag PID of \textcolor{black}{$I(Y(t_n);\textbf{X}(t_n))$} presupposes to work with memoryless processes, a condition that is typically not satisfied in practice as the presence of a temporal statistical structure is inherently expected in time series data. On the other hand, the PID of the TE \textcolor{black}{$I(Y(t_n);\textbf{X}(t_{<n})|Y(t_{<n}))$} cannot account for instantaneous interactions among the processes, nor for causal interactions occurring in the causal direction from the target to the sources.
To overcome these limitations, in the following we propose a framework for the decomposition of the information shared dynamically between the target and the source processes.

\subsection{Partial Information Rate Decomposition}

The framework proposed here for the decomposition of information in multivariate random processes makes use of the concepts of \textit{entropy rate} (ER) and \textit{mutual information rate} (MIR). The ER of a generic stationary \textcolor{black}{vector} random processes $\textcolor{black}{\textbf{X}}$ is defined as  \cite{cover1999elements}
\begin{equation}
   H_{\textcolor{black}{\textbf{X}}}=\lim_{m \to \infty}\frac{1}{m} H \big( \textcolor{black}{\textbf{X}(t_{n:n+m})} \big),
   \label{ER}
\end{equation}
\textcolor{black}{where $\textbf{X}(t_{n:n+m})=\{\textbf{X}(t_n),\ldots, \textbf{X}(t_{n+m}) \}$}.
Being equivalent to the conditional entropy of the present state of the process given its past states (i.e., \textcolor{black}{$H_{\textbf{X}}=H(\textbf{X}(t_n)|\textbf{X}(t_{<n}))$}), the ER quantifies the rate of generation of new information in the process. Then, considering another process $Y$, the MIR between $\textcolor{black}{\textbf{X}}$ and $Y$ is defined as \cite{duncan1970calculation}
\begin{equation}
   I_{\textcolor{black}{\textbf{X}};Y}=\lim_{m \to \infty} \frac{1}{m} I \big( \textcolor{black}{\textbf{X}(t_{n:n+m});Y(t_{n:n+m})} \big),
   \label{MIR}
\end{equation}
quantifying the information shared by the two processes per unit of time; the MIR can be expressed in terms of entropy rates as $I_{\textcolor{black}{\textbf{X}};Y}=H_{\textcolor{black}{\textbf{X}}}+H_{Y}-H_{\textcolor{black}{\textbf{X}},Y}$, evidencing the analogy between the concepts of entropy and MI for random variables and the concepts of ER and MIR for random processes.

Exploiting this analogy, we use the MIR as a building block for assessing and decomposing the dynamic information shared between the scalar target process $Y$ and the vector source process $\textcolor{black}{\textbf{X}}=\{X_1,\ldots,X_M\}$ of the analyzed network system \textcolor{black}{(Fig. \ref{fig:intro}c)}. Specifically, we formalize a so-called \textit{Partial Information Rate Decomposition} (PIRD) which makes use of the same lattice structure of the PID \cite{williams2010nonnegative} to expand the MIR between target and sources as:
\begin{equation}
    I_{Y;\textcolor{black}{\textbf{X}}} = \sum_{\alpha \in \mathcal{A}} I^{\delta}_{Y;\textcolor{black}{\textbf{X}}_{\alpha}},
    \label{PIRD_atoms}
\end{equation}
where $I^{\delta}_{\cdot;\cdot}$ is a PI rate function defined for each atom $\alpha=\{\alpha_1,\ldots,\alpha_J\}$ of the lattice, and $\textcolor{black}{\textbf{X}}_{\alpha}$ denotes the $\alpha^{\mathrm{th}}$ set of subsets of source processes, with $\textcolor{black}{\textbf{X}}_{\alpha_j} \subseteq \textcolor{black}{\textbf{X}}$.
As happens for the PID, to solve the PIRD it is necessary to define a so-called \textit{redundancy rate} function, here denoted as $I^{\cap}_{\cdot;\cdot}$, which generalizes the MIR over the lattice and replaces the concept of redundancy function generalizing the MI. The redundancy rate of the $\alpha^{\mathrm{th}}$ atom is obtained summing the PI rate of the same atom to the PI rates of the atoms positioned downwards in the lattice:  
\begin{equation}
    I^{\cap}_{Y;\textcolor{black}{\textbf{X}}_{\alpha}}=\sum_{\beta \preceq \alpha} I^{\delta}_{Y;\textcolor{black}{\textbf{X}}_{\beta}};
    \label{redundancy_rate_function}
\end{equation} 
then, once the redundancy rate is known, the information rate associated to all atoms can be retrieved via  M\"{o}bius inversion of  (\ref{redundancy_rate_function}). In the Section "Frequency-domain PIRD" we will elaborate on the definition of redundancy rate functions.

Importantly, since the PIRD is built over the same lattice backbone as the PID, several concepts and relations defined for the PID still hold in the dynamic case (see Fig. \ref{fig:fig_lattice}, with the shrewdness of considering information rates in place of the static information distributed over the lattice).
For instance, as the PIRD satisfies the same consistency equations valid for the PID, the redundancy function computed for an atom composed by one single source reduces to the MIR between the target and that source, i.e. $I^{\cap}_{Y;\textcolor{black}{\textbf{X}}_{\alpha}}=I_{Y;X_i}$ when $\alpha=\{i\}$.
Moreover, the rate of information shared between the $M$ source processes $X_1,\ldots,X_M$ taken together and the target process $Y$ can be expanded in analogy to (\ref{PID_contributions}) as the sum of $M+2$ contributions: 
\begin{equation}
    I_{Y;\textcolor{black}{\textbf{X}}} =  \sum_{m=1}^{M} \mathcal{U}_{Y;X_m} + \mathcal{R}_{Y;\textcolor{black}{\textbf{X}}} + \mathcal{S}_{Y;\textcolor{black}{\textbf{X}}},
    \label{PIRD_contributions}
\end{equation}
achieving a so-called \textit{coarse-grained} PIRD whereby each of the $M$ terms $\mathcal{U}_{Y;X_m} = I_{Y;X_m} - \mathcal{R}_{Y;\textcolor{black}{\textbf{X}}}$ identifies the \textit{unique} rate of information produced dynamically by $Y$ that is shared exclusively with $X_m$ ($m=1,\ldots,M$), the term $\mathcal{R}_{Y;\textcolor{black}{\textbf{X}}}$ identifies the \textit{redundant} rate of information that is shared simultaneously with all the source processes in \textcolor{black}{$\textbf{X}$}, and the term $\mathcal{S}_{Y;\textcolor{black}{\textbf{X}}}$ identifies the \textit{synergistic} rate of information that only arises from knowing all the source processes $X_1,\ldots,X_M$.
\textcolor{black}{Moreover, a quantity  reflecting the dynamic redundancy-synergy balance can be obtained as $\Delta_{Y;\textbf{X}}=\mathcal{R}_{Y;\textbf{X}}-\mathcal{S}_{Y;\textbf{X}}$.}
These coarse-grained information rates correspond for the case of two sources to the PI rate of the four atoms of the redundancy rate lattice (Fig. \ref{fig:fig_lattice}a, right), while they can be obtained for the case $M=3$ by summing the PI rates of the atoms with redundant, unique or synergistic character (respectively, magenta, grey and blue shades in Fig. \ref{fig:fig_lattice}b, right).

As a further remark, we stress that the PIRD formulated in (\ref{PIRD_atoms}) decomposes the information rates shared by multivariate processes accounting for their full dynamical structure and, as such, it generalizes previous attempts to apply the PID to random processes \cite{kay2022comparison,varley2023information,koccillari2023behavioural,luppi2024information}. This aspect becomes apparent considering that the MIR (\ref{MIR}) can be expanded as the sum of three terms, two related to the causal (time-lagged) information transfer along the two directions $\textcolor{black}{\textbf{X}} \rightarrow Y$ and $Y \rightarrow \textcolor{black}{\textbf{X}}$, and the latter related to the instantaneous (zero-lag) information shared between \textcolor{black}{$\textbf{X}$} and $Y$ (see, e.g., \cite{bara2023comparison}):
\begin{equation}
   I_{\textcolor{black}{\textbf{X}};Y}=T_{\textcolor{black}{\textbf{X}}\rightarrow Y}+T_{Y\rightarrow \textcolor{black}{\textbf{X}}}+I_{\textcolor{black}{\textbf{X}} \cdot Y},
   \label{MIR_dec}
\end{equation}
where \textcolor{black}{$T_{\textbf{X}\rightarrow Y}=I(Y(t_n);\textbf{X}(t_{<n})|Y(t_{<n}))$ and $T_{Y\rightarrow \textbf{X}}=I(\textbf{X}(t_n);Y(t_{<n})|\textbf{X}(t_{<n}))$} are the transfer entropies measuring causal information transfer, and \textcolor{black}{$I_{\textbf{X} \cdot Y}=I(\textbf{X}(t_n);Y(t_n)|\textbf{X}(t_{<n}),Y(t_{<n}))$} is the instantaneous information shared between the processes.
Now, if we consider the case in which the overall multivariate process $\textcolor{black}{\textbf{Z}}=\{\textcolor{black}{\textbf{X}},Y\}$ is memoryless, meaning that its whole temporal statistical structure is absent, the two TEs vanish and the instantaneous term becomes the information shared instantaneously between \textcolor{black}{$\textbf{X}$} and $Y$, showing how the PIRD reduces to a PID applied to the static MI \textcolor{black}{$I(\textbf{X}(t_n);Y(t_n))$}. If, on the other hand, we consider the case of strictly causal processes with exclusive (unidirectional) information transfer from the sources to the target, both the instantaneous term and the TE from $Y$ to \textcolor{black}{$\textbf{X}$} vanish, and the PIRD reduces to a PID applied to the TE $T_{\textcolor{black}{\textbf{X}}\rightarrow Y}$. These properties will be illustrated in simulated examples in Sect. "Theoretical Examples".

\subsection{\label{subsect:SpectralPIRD}Frequency-domain PIRD}
To solve the PIRD identified by (\ref{PIRD_atoms}), it is necessary to define a redundancy rate function taking values over the lattice underlying the decomposition. In principle, the redundancy rate can be defined following any of the several approaches formulated for the PID, adapting it to the calculation of the MIR between random processes in place of the MI between random variables.
In the PID literature, after the seminal work by Williams and Beer who first recognized that the concept of redundancy cannot be defined using classical information theory and proposed an axiomatic definition of redundant information \cite{williams2010nonnegative}, a number of alternative redundancy measures adopting or modifying the original set of axioms have been proposed \cite{bertschinger2014quantifying, barrett2015exploration, ince2017measuring, pakman2021estimating,makkeh2021introducing, ehrlich2024partial}.
\textcolor{black}{Among them, a simple definition of redundancy is the MMI measure \cite{barrett2015exploration}, which can be straightforwardly adapted to information rates; specifically, a so-called MMI redundancy rate can be defined taking the minimum MIR over the source processes composing the analyzed atom $\alpha=\{\alpha_1,\ldots,\alpha_J \}$ of the redundancy lattice:
\textcolor{black}{
\begin{equation}
    I^{\cap}_{Y;\textbf{X}_{\alpha}} = I^{\mathrm{MMI}}_{Y;\textbf{X}_{\alpha}} := \min\limits_{j=1,\ldots,J} I_{Y;\textbf{X}_{\alpha_j}}.
    \label{redundancy_MMI}
\end{equation}
}
As an alternative to using directly (\ref{redundancy_MMI}), in this work we exploit the MMI principle at the pointwise level in the spectral domain.
Traditional pointwise methods applied to multiple random variables} derive "local" redundancy measures working on the specific realizations of the variables at hand, rather than on the variable themselves, and then compute "global" redundancy via statistical expectation, i.e. taking the ensemble average over all possible realizations \cite{ince2017measuring,finn2018pointwise,makkeh2021introducing,gutknecht2021bits}.
The same approach can be followed for the PIRD, e.g., working on the local version of the MIR (\ref{MIR}) to formalize the notion of pointwise redundancy rate; the redundancy rate among processes could be then retrieved by ensemble averaging, which for stationary processes corresponds to time-domain averaging. 
Here, we propose an approach that is conceptually similar, but is implemented through a pointwise representation in frequency rather than in time.
Specifically, we characterize the analyzed network of random processes in the frequency domain, considering the information provided about a particular oscillatory component of the target process $Y$ by the iso-frequency oscillatory components of the source processes collected in \textcolor{black}{$\textbf{X}$}. The idea is to perform the entire PIRD on the pointwise level for a particular frequency, i.e., to decompose the spectral (frequency-specific) MIR denoted as $i_{Y;\textcolor{black}{\textbf{X}}}(\omega)$, where $\omega=2\pi \frac{f}{f_s}$ is the normalized circular frequency ($\omega \in [-\pi, \pi]$, $f \in [-\frac{f_s}{2},\frac{f_s}{2}]$); the spectral MIR is identified from the expansion of the MIR in the frequency domain as:
\begin{equation}
    I_{Y;\textcolor{black}{\textbf{X}}}=\frac{1}{2\pi}\int_{-\pi}^{\pi}i_{Y;\textcolor{black}{\textbf{X}}}(\omega) \mathrm{d}\omega.
    \label{spectralMIR_integral}
\end{equation}
While different integral transforms (e.g., the wavelet transform) could in principle be used to expand the MIR, in the next section we will use a definition of spectral MIR which satisfies (\ref{spectralMIR_integral}) for Gaussian processes to formalize the PIRD in the frequency domain. 
Crucially, (\ref{spectralMIR_integral}) connects the time- and frequency-domain representations of information-theoretic quantities for random processes, and is exploited here to relate the \textcolor{black}{time-domain} PIRD (\ref{PIRD_atoms}) to its frequency-domain extension. Such an extension is denoted as \textit{spectral} PIRD and is formulated, for the oscillatory components of the network process assessed at the frequency $\omega$, expressing the spectral MIR as:
\begin{equation}
    i_{Y;\textcolor{black}{\textbf{X}}}(\omega) = \sum_{\alpha \in \mathcal{A}} i^{\delta}_{Y;\textcolor{black}{\textbf{X}}_{\alpha}}(\omega),
    \label{Spectral_PIRD_atoms}
\end{equation}
where $i^{\delta}_{\cdot;\cdot}(\omega)$ is the \textit{spectral partial information rate} function defined over a lattice specifically identified on the oscillations with frequency $\omega$, and $\textcolor{black}{\textbf{X}}_{\alpha}$ denotes the set of subsets of source processes indexed by the atom $\alpha$. 
As happens with PID, to solve (\ref{Spectral_PIRD_atoms}) we need to identify a \textit{spectral redundancy rate}, $i^{\cap}_{Y;\textcolor{black}{\textbf{X}}_{\alpha}}(\omega)$, generalizing the spectral MIR over the lattice, and express it as the sum of the spectral PI rates of the atoms positioned downwards in the lattice:
\begin{equation}
    i^{\cap}_{Y;\textcolor{black}{\textbf{X}}_{\alpha}}(\omega) = \sum_{\beta \preceq \alpha} i^{\delta}_{Y;\textcolor{black}{\textbf{X}}_{\beta}}(\omega).
    \label{red_from_PI_spectral}
\end{equation}
From (\ref{red_from_PI_spectral}), the spectral PI rate can be computed recursively, in analogy to (\ref{delta_from_red}), as:
\begin{equation}
    i^{\delta}_{Y;\textcolor{black}{\textbf{X}}_{\alpha}}(\omega) = i^{\cap}_{Y;\textcolor{black}{\textbf{X}}_{\alpha}}(\omega) - \sum_{\beta \prec \alpha} i^{\delta}_{Y;\textcolor{black}{\textbf{X}}_{\beta}}(\omega).
    \label{atom_information_rate_function_PIRD}
\end{equation}
Furthermore, once the atoms are identified via (\ref{atom_information_rate_function_PIRD}), they can be properly grouped to obtain a coarse-grained representation of the spectral PIRD which takes the form:
\begin{equation}
    i_{Y;\textcolor{black}{\textbf{X}}}(\omega) =  \sum_{m=1}^{M} u_{Y;X_m}(\omega) + r_{Y;\textcolor{black}{\textbf{X}}}(\omega) + s_{Y;\textcolor{black}{\textbf{X}}}(\omega),
    \label{spectralPIRD_contributions}
\end{equation}
where the $M+2$ atoms reflect the $M$ unique contributions of each source process, as well as the redundant and synergistic contributions of all sources, to the rate of information produced by the target process at the specific frequency $\omega$; the coarse-grained spectral PIRD is obtained adopting the same criteria already described for the PID \cite{rosas2020reconciling}, leading to the coarse-grained atoms illustrated in Fig. \ref{fig:fig_lattice} (right panels).

Eq. (\ref{atom_information_rate_function_PIRD}) provides a solution for the spectral PIRD up to the definition of a proper spectral redundancy rate function.
Here, we propose to assess spectral redundancy following the MMI principle \cite{barrett2015exploration} applied to the spectral MIR computed between the target and each source process at the frequency of interest. Specifically, 
we define the frequency-specific redundancy rate function of the atom $\alpha=\{\alpha_1,\ldots,\alpha_J \}$ of the spectral redundancy lattice, \textcolor{black}{denoted as spectral minimum mutual information (SMMI) redundancy rate}, as follows:
\begin{equation}
    \textcolor{black}{i^{\cap}_{Y;\textbf{X}_{\alpha}}(\omega)=i^{\mathrm{SMMI}}_{Y;\textbf{X}_{\alpha}}(\omega) := \min\limits_{j=1,\ldots,J} i_{Y;\textbf{X}_{\alpha_j}}(\omega),}
    \label{spectral_redundancy_rate_function_MMI}
\end{equation}
where $J=|\alpha|$ indicates the cardinality of the atom; e.g., $J=1$ if $\alpha = \{1\}$ or $\alpha = \{12\}$, while $J=2$ if $\alpha = \{\{3\},\{12\}\}$. 
Note that, in contrast with the classical MMI formulation performed for random variables \cite{barrett2015exploration}, here the minimum MIR is searched at the pointwise (frequency-specific) level. While a pointwise implementation of the MMI criterion would be cumbersome if performed in the time domain, and it is indeed avoided by the existing pointwise approaches  \cite{ince2017measuring,finn2018pointwise,makkeh2021introducing} because the local MI or the local MIR can take negative values, in our case is favored by the non-negativity of the spectral MIR. The properties of the spectral redundancy rate (\ref{spectral_redundancy_rate_function_MMI}) computed for Gaussian processes will be discussed in the following subsection \textcolor{black}{and in Appendix A}.

\textcolor{black}{Eq. (\ref{spectral_redundancy_rate_function_MMI}) defines the redundancy rate and the related PIRD in the frequency domain. Then, to quantify the redundancy rate in the time domain we exploit the idea pursued by pointwise approaches of averaging local redundancy measures to get global measures  \cite{ince2017measuring,finn2018pointwise,makkeh2021introducing,gutknecht2021bits}, with the difference that averaging is performed over frequency rather than over time. Specifically, the time-domain (global) version of the SMMI redundancy rate is defined as the normalized full-frequency integral of the spectral redundancy:
\begin{equation}
I^{\cap}_{Y;\textbf{X}_{\alpha}}=I^{\mathrm{SMMI}}_{Y;\textbf{X}_{\alpha}}:=\frac{1}{2\pi}\int_{-\pi}^{\pi} i^{\mathrm{SMMI}}_{Y;\textbf{X}_{\alpha}}(\omega) \mathrm{d}\omega. 
    \label{time_redundancy_rate_SMMI}
\end{equation}
The redundancy rate defined in (\ref{time_redundancy_rate_SMMI}) is finally used to derive the PI rate associated to the $\alpha^{\mathrm{th}}$ atom of the lattice in analogy to (\ref{delta_from_red}) as:
\begin{equation}
    I^{\delta}_{Y;\textbf{X}_{\alpha}} = I^{\cap}_{Y;\textbf{X}_{\alpha}} - \sum_{\beta \prec \alpha} I^{\delta}_{Y;\textbf{X}_{\beta}},
    \label{atom_information_rate_time_PIRD}
\end{equation}
or in a compact way via  M\"{o}bius inversion of  (\ref{redundancy_rate_function}), thus completing the PIRD. Moreover, the PI rates of the various atoms can be aggregated as in (\ref{PIRD_contributions}) to yield a coarse-grained PIRD.
Importantly, the SMMI redundancy rate defined in (\ref{time_redundancy_rate_SMMI}) is different than the MMI redundancy rate given in (\ref{redundancy_MMI}); in Appendix A.1 we demonstrate that $I^{\mathrm{SMMI}}_{Y;\textbf{X}_{\alpha}} \leq I^{\mathrm{MMI}}_{Y;\textbf{X}_{\alpha}}$, with equivalence holding in the absence of temporal correlations (i.e. when the processes reduce to i.i.d. random variables). This discrepancy occurs because the MMI definition is not additive in the sense that averaging over frequencies the minimum spectral MIR does not yield the minimum time-domain MIR. Nevertheless, we take this as an advantage of the definition of redundancy rate (\ref{time_redundancy_rate_SMMI}),} \textcolor{black}{as a less conservative measure of redundancy can help reducing the overestimation of redundancy in which definitions based on the minimum MI principle may incur \cite{timme2014synergy, griffith2015quantifying,harder2013bivariate,ince2017measuring}.}

From the perspective of information decomposition based on lattice structures \cite{williams2010nonnegative}, the PID 
(\ref{PID_atoms},\ref{redundancy_function}) is conceptually equivalent
to the PIRD (\ref{PIRD_atoms},\ref{redundancy_rate_function}) and to the spectral PIRD (\ref{Spectral_PIRD_atoms},\ref{red_from_PI_spectral}); what changes in the three formulations is only the quantity to be decomposed, from the MI $I(T;\textcolor{black}{\textbf{S}})$ for the case of random variables to the MIR $I_{Y;\textcolor{black}{\textbf{X}}}$ for the case of random processes, and to the spectral MIR $i_{Y;\textcolor{black}{\textbf{X}}}(\omega)$ for the case of oscillatory components of random processes. Since these quantities maintain the same meaning and properties (i.e., they are non-negative measures of shared information), the three formulations of PID, PIRD and spectral PIRD will lead to a unique solution for the atoms once a redundancy function (respectively, the redundancy among collections of random variables, the redundancy rate among collections of random processes, and the spectral redundancy rate among collections of oscillatory components of random processes) is fixed over a lattice structure like that in Fig. \ref{fig:fig_lattice}.
\textcolor{black}{Moreover}, moving from the PIRD defined on the pointwise level (frequency-specific) to that defined on the process-level (time domain) is straightforward exploiting spectral integration.
\textcolor{black}{This is guaranteed by the property of linearity of the definite integrals, which allows e.g. to obtain (\ref{atom_information_rate_time_PIRD}) through full-frequency integration of (\ref{atom_information_rate_function_PIRD}).}
In fact, solving the frequency-specific PIRD to obtain the PI rate $i^{\delta}_{Y;\textcolor{black}{\textbf{X}}_{\alpha}}(\omega)$ for each atom $\alpha$ via (\ref{atom_information_rate_function_PIRD}) and then integrating these contributions along the whole frequency axis to get time domain values is equivalent to integrating the spectral redundancy rate functions $i^{\cap}_{Y;\textcolor{black}{\textbf{X}}_{\alpha}}(\omega)$ in the range $[ -\pi, \pi]$ via (\ref{time_redundancy_rate_SMMI}) and then applying (\ref{atom_information_rate_time_PIRD}) in the time domain to get the contributions $I^{\delta}_{Y;\textcolor{black}{\textbf{X}}_{\alpha}}$ \textcolor{black}{(proof in the Appendix A.2)}.
\textcolor{black}{The tight link between the spectral and time-domain formulations of the proposed PIRD schemes allows also the useful possibility of tailoring the information decomposition to predetermined frequency bands with practical meaning. Indeed, frequency-specific PIRD schemes can be designed} whereby the decomposition of multivariate information rates is achieved in the time domain but is limited to oscillatory components whose frequencies are confined within a specific band of the spectrum. This band-limited decomposition, which is useful to highlight and decompose multivariate interactions with focus on specific oscillatory components of the analyzed processes, will be illustrated in the next sections in both simulated and applicative settings.

\subsection{\label{subsect_PIRD_processes_linearformulation}Formulation for Gaussian processes}
In the linear signal processing framework, the analyzed set of stochastic processes $\textcolor{black}{\textbf{Z}}=\{Y,X_1,\ldots,X_M\}$ can be described in terms of its power spectral density (PSD) matrix expressed as
\begin{equation}
    \textbf{P}_{\textcolor{black}{\textbf{Z}}}(\omega)=\begin{bmatrix}
    P_{Y}(\omega) & P_{YX_1}(\omega) & \cdots & P_{YX_M}(\omega)\\
    P_{X_1Y}(\omega) & P_{X_1}(\omega) & \cdots & P_{X_1X_M}(\omega)\\
    \vdots & \vdots & \ddots & \vdots\\
    P_{X_MY}(\omega) & P_{X_MX_1}(\omega) & \cdots & P_{X_M}(\omega)\\    
    \end{bmatrix},
    \label{PSDmatrixS}
\end{equation}
which contains the individual PSDs of the processes $\{ Y, X_1, \ldots, X_M \}$ as diagonal elements and the cross-PSDs between them as off-diagonal elements. Individual and cross-PSDs are defined as the Fourier Transform (FT) of the covariance functions of the processes: i.e., $P_{Y}(\omega)=\mathfrak{F}\{R_{Y}(k)\}$\textcolor{black}{, $P_{X_iX_j}(\omega)=\mathfrak{F}\{R_{X_iX_j}(k)\}$} and $P_{YX_i}(\omega)=\mathfrak{F}\{R_{YX_i}(k)\}$, \textcolor{black}{where $R_{Y}(k)=\mathbb{E}[Y(t_n)Y(t_{n-k})]$, $R_{X_iX_j}(k)=\mathbb{E}[X_{i}(t_n)X_{j}(t_{n-k})]$ and $R_{YX_i}(k)=\mathbb{E}[Y(t_n)X_{i}(t_{n-k})]$} are the auto-correlation function of $Y$\textcolor{black}{, the cross-correlation function between $X_i$ and $X_j$} and the cross-correlation function between $Y$ and $X_i$, respectively ($k$ is the correlation delay); correlations are equivalent to covariances for zero-mean processes.

The PSD matrix is the central element for the implementation of the spectral PIRD. In fact, it is well-known that, for jointly Gaussian processes \textcolor{black}{$\textbf{X}$} and $Y$, the MIR admits the following spectral expansion \cite{gelfand1959calculation, chicharro2011spectral}:
\begin{equation}
    I_{\textcolor{black}{\textbf{X}};Y}=\frac{1}{4\pi}\int_{-\pi}^{\pi} \log\frac{|\textbf{P}_{\textcolor{black}{\textbf{X}}}(\omega)| P_{Y}(\omega)}{|\textbf{P}_{\textcolor{black}{\textbf{Z}}}(\omega)|}\mathrm{d}\omega,
    \label{spectralMIR_expansion}
\end{equation}
where $\textbf{P}_{\textcolor{black}{\textbf{X}}}(\omega)$ is the PSD matrix pruned of the first row and column, and $|\cdot|$ denotes matrix determinant. The spectral MIR decomposed by the spectral PIRD then results immediately comparing (\ref{spectralMIR_expansion}) with (\ref{spectralMIR_integral}):
\begin{equation}
    i_{Y;\textcolor{black}{\textbf{X}}}(\omega)=\frac{1}{2}\log \frac{ |\textbf{P}_{\textcolor{black}{\textbf{X}}}(\omega)| P_{Y}(\omega)}{|\textbf{P}_{\textcolor{black}{\textbf{Z}}}(\omega)|}.
    \label{spectralMIR}
\end{equation}
Moreover, for any given atom $\alpha=\{\alpha_1,\ldots,\alpha_J \}$ of the spectral redundancy lattice, the application of (\ref{spectralMIR}) is particularized to the element $\alpha_j$ as follows:
\begin{equation}
    i_{Y;\textcolor{black}{\textbf{X}}_{\alpha_j}}(\omega)=\frac{1}{2}\log \frac{|\textbf{P}_{\textcolor{black}{\textbf{X}}_{\alpha_j}}(\omega)| P_{Y}(\omega) }{|\textbf{P}_{[Y\textcolor{black}{\textbf{X}}_{\alpha_j}]}(\omega)|},
    \label{spectralMIR_alphaj}
\end{equation}
where $\textbf{P}_{[Y\textcolor{black}{\textbf{X}}_{\alpha_j}]}(\omega)$ is a sub-matrix identified from $\textbf{P}_{\textcolor{black}{\textbf{Z}}}(\omega)$ as
\begin{equation}
    \textbf{P}_{[Y\textcolor{black}{\textbf{X}}_{\alpha_j}]}(\omega)=\begin{bmatrix}
    P_{Y}(\omega) & \textbf{P}_{Y\textcolor{black}{\textbf{X}}_{\alpha_j}}(\omega)\\
    \textbf{P}_{\textcolor{black}{\textbf{X}}_{\alpha_j}Y}(\omega) & \textbf{P}_{\textcolor{black}{\textbf{X}}_{\alpha_j}}(\omega)\\
    \end{bmatrix};
    \label{PSDmatrixYXalphaj}
\end{equation}
the computation of (\ref{spectralMIR_alphaj}) for each $j=1,\ldots, J$ yields the spectral MIR
terms to be used in (\ref{spectral_redundancy_rate_function_MMI}) for the computation of \textcolor{black}{the SMMI spectral redundancy rate}.

The SMMI spectral redundancy rate defined by (\ref{spectral_redundancy_rate_function_MMI},\ref{spectralMIR_alphaj}) is a proper redundancy function as it satisfies the axioms originally proposed by Williams and Beer \cite{williams2010nonnegative}, i.e., \textit{\textcolor{black}{weak} symmetry}: 
$i^{\cap}_{Y;\textcolor{black}{\textbf{X}}_{\{\alpha_1,\ldots,\alpha_J\}}}(\omega)$ is symmetric w.r.t. the $\alpha_j, j=1,\ldots,J$; \textit{self-redundancy}: $i^{\cap}_{Y;\textcolor{black}{\textbf{X}}_{\alpha}}(\omega)=i_{Y;\textbf{X}_{\alpha}}(\omega)$ when $|\alpha|=1$; \textit{monotonicity}: $i^{\cap}_{Y;\textcolor{black}{\textbf{X}}_{\{\alpha_1,\ldots,\alpha_{J-1},\alpha_J\}}}(\omega) \leq i^{\cap}_{Y;\textcolor{black}{\textbf{X}}_{\{\alpha_1,\ldots,\alpha_{J-1}\}}}(\omega)$, with equality if $\alpha_{J-1}\subseteq \alpha_J$; \textcolor{black}{and \textit{non-negativity}: $i^{\cap}_{Y;\textbf{X}_{\alpha}}(\omega) \geq 0$.}
These properties, which are derived from the minimum MIR definition  (\ref{spectral_redundancy_rate_function_MMI}) of the spectral redundancy rate, and from the properties of non-negativity and monotonicity of the spectral MIR (\ref{spectralMIR_alphaj}) \cite{nedungadi2011block}, \textcolor{black}{hold equivalently for the spectral and time-domain PIRD (proofs are in Appendix A.3).} \textcolor{black}{Nevertheless, the SMMI measure of redundancy rate based on (\ref{spectralMIR_alphaj}) suffers from important limitations, the main being that it is valid only for stationary Gaussian processes admitting a spectral representation \cite{chicharro2011spectral}, and it lacks additivity \cite{rauh2023continuity} in the sense that the minimum MIR of the collection of independent sets of source and target processes is not equal to the sum of the minimum MIRs evaluated separately for each set.}

We conclude discussing the practical computation of the PSD matrix (\ref{PSDmatrixS}) whose elements are exploited to estimate all the MIR terms entering the PIRD.
There are several standard methods for computing the PSD matrix of multivariate time-series, including the averaged periodogram, multi taper techniques, and wavelet approaches \cite{von2020nonparametric}. Here, we use the computation that induces a linear parametric representation of the observed dynamics. Specifically, the analyzed stochastic process $\textcolor{black}{\textbf{Z}}=\{Y,X_1,\ldots,X_M\}$ is described as a vector autoregressive (VAR) process \cite{lutkepohl2005new}:
\begin{equation}
   \textcolor{black}{\textbf{Z}(t_n)} = \sum_{k=1}^{p}\mathbf{A}_k \textcolor{black}{\textbf{Z}(t_{n-k})} + \textcolor{black}{\textbf{U}(t_n)}
   \label{VAR_model},
\end{equation}
where $p$ is the model order, defining the maximum lag used to quantify interactions, \textcolor{black}{$\textbf{Z}(t_n)=[Y(t_n) X_{1}(t_n) \ldots X_{M}(t_n)]^\intercal$} is a $M+1$-dimensional vector collecting the present state of all processes, $\mathbf{A}_k$ is the $(M+1) \times (M+1)$ matrix of the model coefficients relating the present with the past of the processes at lag $k$, and \textcolor{black}{$\textbf{U}(t_n)=[U_{Y}(t_n) U_{X_1}(t_n) \ldots U_{X_M}(t_n)]^\intercal$} is a vector of $M+1$ zero-mean white innovations with $(M+1) \times (M+1)$ positive definite covariance matrix \textcolor{black}{$\mathbf{\Sigma}_\textbf{U}=\mathbb{E}[\textbf{U}(t_n) \textbf{U}(t_n)^\intercal]$}.
To analyze the VAR model (\ref{VAR_model}) in the frequency domain, the FT of (\ref{VAR_model}) is taken to derive
\begin{equation}
   \textcolor{black}{\textbf{Z}}(\omega) = [\mathbf{I} - \sum_{k=1}^{p}\mathbf{A}_k e^{-\mathbf{j} \omega k}]^{-1} \textcolor{black}{\textbf{U}}(\omega)=\mathbf{H}(\omega)\textcolor{black}{\textbf{U}}(\omega)
   \label{VAR_model_freq},
\end{equation}
where $\textcolor{black}{\textbf{Z}}(\omega)$ and $\textcolor{black}{\textbf{U}}(\omega)$ are the FTs of \textcolor{black}{$\textbf{Z}(t_n)$} and \textcolor{black}{$\textbf{U}(t_n)$}, $\mathbf{j}=\sqrt{-1}$ and $\mathbf{I}$ is the $(M+1)$-dimensional identity matrix. The $(M+1) \times (M+1)$ matrix $\mathbf{H}(\omega)$ contains the transfer functions relating the FTs of the innovation processes in \textcolor{black}{$\textbf{U}$} to the FTs of the processes in \textcolor{black}{$\textbf{Z}$}. Finally, the transfer matrix is exploited, together with the covariance of the VAR innovations, to derive the PSD matrix through spectral factorization \cite{wilson1972factorization}:
\begin{equation}
    \textbf{P}_{\textcolor{black}{\textbf{Z}}}(\omega)=\mathbf{H}(\omega)\mathbf{\Sigma}_{\textcolor{black}{\textbf{U}}}\mathbf{H}^*(\omega)
   \label{SpectralFactorization},
\end{equation}
where $^*$ stands for conjugate transpose. 

The practical computation of the PSD starts from the  parameters $\textbf{A}_k$ and $\mathbf{\Sigma}_{\textcolor{black}{\textbf{U}}}$ of the VAR model (\ref{VAR_model}), which can be easily estimated from realizations of the process \textcolor{black}{$\textbf{X}$} available in the form of multivariate time series. While several procedures exist to perform VAR model identification \cite{barnett2014mvgc,antonacci2020information}, here we use a tool \cite{faes2012measuring} based on classical least squares estimation \cite{lutkepohl2005new} implemented optimizing the model order $p$ through the Akaike information criterion \cite{akaike1974new}; 
remarkably, high model orders denote a non-parsimonious representation of the multivariate time series at hand, which however may indicate that the linear representation induced by the VAR model can capture complex, potentially nonlinear dynamics of the underlying processes, provided that the fitted VAR model is stable \cite{barnett2015granger}. 

In closing this section we note that alternative approaches for the computation of multivariate information measures, which are not dealt with here, can rely on the model-free computation of entropy measures, which guarantees high flexibility in the representation of the dynamics but also exposes to issues related to the data-efficient estimation of entropy measures \cite{runge2012escaping,vicente2011transfer,xiong2017entropy,ricci2021estimating}.
\textcolor{black}{In contrast, the approach formulated in this section based on linear spectral analysis offers high computational reliability and thus favors the practical use of PIRD in a big variety of network systems encountered in real-world applications. Moreover, it is worth stressing that fitting a VAR model for computing the information decomposition measures assumes a linear model rather than a linear process, which is a subtle but important distinction: while the model has a linear structure, this does not necessarily mean that the underlying process must be linear in its dynamics. Furthermore, the presence of non-linearities in the data does not inherently imply that the fitted model will be unstable: in fact, Wold’s decomposition theorem guarantees that any stationary process admits a linear model representation, although this model may require an infinite order \cite{hannan1979statistical}. The critical question remains about whether the linear model applied to time series generated by a nonlinear process provides a sufficiently accurate representation of the process dynamics, and this relates to the model order that may be potentially infinite \cite{barnett2015granger,barnett2023dynamical}. In general, choosing between the linear representation which is simpler but can be non parsimonious and a model-free approach which is theoretically well-posed but very often data demanding remains an open and interesting area for future research.}
In the following sections, keeping the linear model-based approach, we will first illustrate the PIRD framework in theoretical VAR processes where the parameters are set to simulate controlled behaviors, and then apply the framework on real multivariate time series measured in the context of network physiology.

\section{\label{sect:simulations}Theoretical Examples}

In this section, we characterize the behavior of the measures proposed to dissect the multivariate information shared by random processes using linear VAR models; this allows for a theoretical analysis whereby the exact profiles of the measures are computed starting from the true values imposed for the model parameters. Such parameters are varied in different simulation settings to study the time-domain and spectral behavior of the PIRD measures and to compare them with different implementations of the standard PID applied to random processes. 
\textcolor{black}{Simulations are designed to induce expected behaviors and thus aim to provide a validation of the PIRD framework in comparison to standard PID schemes.}

\subsection{\label{subsect:simulations_2sources} Effects of temporal correlations}
\begin{figure*}
    \centering
    \includegraphics[width=\textwidth]{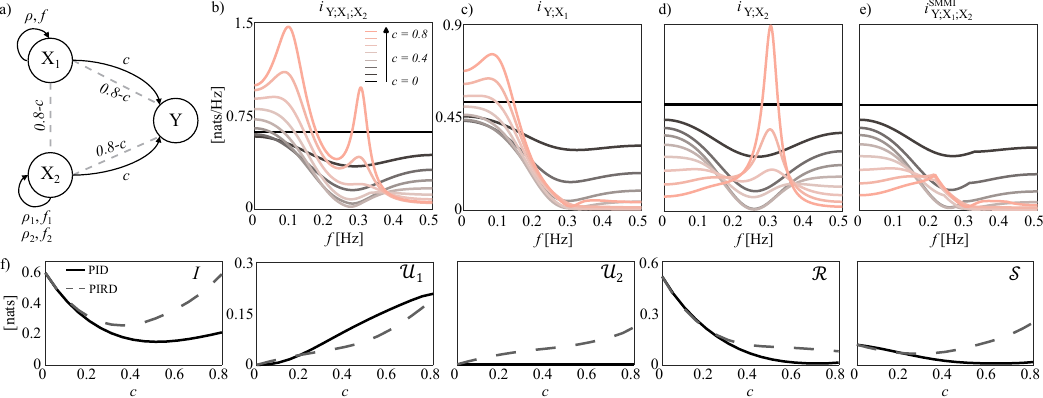}
    \caption{\textbf{The presence of temporal correlations has a profound impact on the multivariate information shared at lag zero by multiple random processes.} \textbf{(a)} Simulation design, where $Y$ is the target process and $\{X_1,X_2\}$ is the group of sources; time-lagged interactions (solid black arrows) and zero-lag interactions (dashed grey lines) are set respectively to increase and decrease with the parameter $c \in [0-0.8]$.
    \textbf{(b-e)} Spectral profiles of the joint MIR between $Y$ and $\{X_1,X_2\}$, of the individual MIR between $Y$ and $X_1$ and between $Y$ and $X_2$, and of the redundancy rate $i^{\textcolor{black}{\mathrm{SMMI}}}_{Y;X_1;X_2}(f)$, obtained at varying the parameter $c$ from zero (continuous black lines) to $0.8$ (continuous pink lines).
    \textbf{(f)} Time-domain behavior of the total information shared between the target and the two sources ($I$), of the unique information shared between the target and each individual source ($\mathcal{U}_1$, $\mathcal{U}_2$), and of the redundant and synergistic information provided by the two sources to the target ($\mathcal{R}$, $\mathcal{S}$), measured using the the zero-lag PID (continuous black lines) and the PIRD (dashed grey lines) as a function of the parameter $c$.}
    \label{fig:thsim_PID_vs_PIRD_spectral_functions}
\end{figure*}

First, we characterize the PIRD in both the time and frequency domains in simulated dynamic networks involving three processes ($\textcolor{black}{\textbf{Z}}=\{Y,X_1,X_2 \}$), comparing the time-domain measures with the standard PID decomposing instantaneous interactions among the processes. To do this we use a three-variate VAR process simulated with $f_s=1$, in which different regimes of dynamic interaction are set by varying the parameters related to zero-lag effects, lagged interactions, and autonomous dynamics. Specifically, the 3-VAR process is defined as:
\begin{equation}
\begin{aligned}
&\textcolor{black}{Y(t_n)} = c \textcolor{black}{X_{1}(t_{n-1})} + c \textcolor{black}{X_{2}(t_{n-2})} + \textcolor{black}{U_{Y}(t_{n})}\\
&\textcolor{black}{X_{1}(t_n)} = \sum_{k=1}^{2} a_{1,k} \textcolor{black}{X_{1}(t_{n-k})} + \textcolor{black}{U_{X_1}(t_{n})} \\
&\textcolor{black}{X_{2}(t_n)} = \sum_{k=1}^{4} a_{2,k} \textcolor{black}{X_{2}(t_{n-k})} + \textcolor{black}{U_{X_2}(t_{n})}
\label{th_sim_3VAR_equations}
\end{aligned}
\end{equation}
where $U_{Y}$, $U_{X_1}$ and $U_{X_2}$ are Gaussian white noises with zero mean and unit variance. The covariance matrix of the residuals,
\begin{equation*}
    \mathbf{\Sigma}_{\textcolor{black}{\textbf{U}}} = \begin{pmatrix}
  1 & \sigma^2_{U_{YX_1}} & \sigma^2_{U_{YX_2}}\\
  \sigma^2_{U_{X_1Y}} & 1 & \sigma^2_{U_{X_1X_2}}\\
  \sigma^2_{U_{X_2Y}} & \sigma^2_{U_{X_2X_1}} & 1
\end{pmatrix},
\end{equation*}
is built in such a way to generate zero-lag cross-correlations among the processes modulated inversely by the parameter $c$, imposing $\sigma^2_{U_{YX_1}} = \sigma^2_{U_{YX_2}} = \sigma^2_{U_{X_1X_2}} = 0.8 - c$. The autonomous oscillations in the two source processes $X_1$ and $X_2$ are obtained placing pairs of complex-conjugate poles, with modulus $\rho$ and phase $2\pi{f}$, in the complex plane representation of each process; the AR coefficients resulting from this setting at lags $1,2$ are $a_1=2\rho \cos(2\pi f)$ and $a_2=-\rho^2$ \cite{faes2015information}. Here, we place a pair of poles for the process $X_1$, setting $\rho=c, f=0.1$ Hz so that the strength of the autonomous dynamics determined by the coefficients $a_{1,1}$ and $a_{1,2}$ depends on the parameter $c$; similarly, we place two pairs of poles for $X_2$, setting $\rho_{1}=c, f_{1}=0.1$ Hz and $\rho_{2}=1.125c, f_{2}=0.3$ Hz so that the strength of the autonomous dynamics of $X_2$ determined by the coefficients $a_{2,k}, k=1,\ldots,4$, depends on the parameter $c$. Moreover, causal interactions are set from $X_1$ to $Y$ at lag $k=1$ and from $X_2$ to $Y$ at lag $k=2$, with strength modulated by the parameter $c$. 
The parameter $c$ is varied in the range $\left[ 0 - 0.8 \right]$, thus allowing (i) progressive strengthening of the autonomous dynamics in the processes $X_1$, $X_2$ and of the causal interaction from $X_1$ and $X_2$ to $Y$, as well as (ii) progressive weakening of the zero-lag interactions among the three processes. The simulation design is shown in Fig. \ref{fig:thsim_PID_vs_PIRD_spectral_functions}a.

\begin{figure}
    \centering
\includegraphics{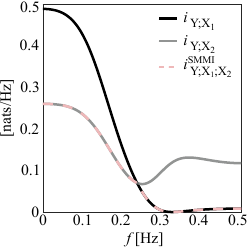}
    \caption{Example of how the spectral redundancy rate function $i^{\cap}_{Y;\textcolor{black}{\textbf{X}}_{\{1\}\{2\}}}(f)=i^{\textcolor{black}{\mathrm{SMMI}}}_{Y;X_1;X_2}(f)$ \textcolor{black}{(dashed pink line)} is computed as the minimum of the interaction between each individual source and the target at the specific frequency $f$, $\min\limits_{i=1,2} i_{Y;X_{i}}(f)$\textcolor{black}{, where the spectral profiles $i_{Y;X_{1}}(f), i_{Y;X_{2}}(f)$ are arbitrarily depicted as solid black and grey lines, respectively}. The spectral PIRD allows to overcome the drawback of the MMI-PID which sets to zero one of the two unique contributions, as well as to delve into the spectral content of the investigated processes and their interactions.}
    \label{fig:thsim_PID_vs_PIRD_redundancy}
\end{figure}

The PIRD was performed computing the \textcolor{black}{SMMI} redundancy rate according to (\ref{spectral_redundancy_rate_function_MMI}), with the spectral MIR function computed as in (\ref{spectralMIR_alphaj}) after deriving the PSD matrix from the VAR parameters, and then obtaining the spectral PI rate through (\ref{atom_information_rate_function_PIRD}) which $-$in this case with $M=2$ sources$-$ yields immediately the coarse-grained terms in (\ref{spectralPIRD_contributions}); all these terms were then integrated over the full frequency axis to obtain the unique $\mathcal{U}_{Y;X_1}$, $\mathcal{U}_{Y;X_2}$, redundant $\mathcal{R}_{Y;X_1;X_2}$ and synergistic $\mathcal{S}_{Y;X_1;X_2}$ information rates in the time domain.
An example of how spectral redundancy is computed as the minimum MIR at each frequency $f$ is shown in Fig. \ref{fig:thsim_PID_vs_PIRD_redundancy} for one specific VAR configuration ($c$ = 0.5). Fig. \ref{fig:thsim_PID_vs_PIRD_spectral_functions} reports the spectral MIR and redundancy rate functions, as well as the time-domain values of the PIRD terms, investigated at varying the simulation parameter $c$. In Fig. \ref{fig:thsim_PID_vs_PIRD_spectral_functions}f the time-domain PIRD is compared with the instantaneous PID which decomposes the MI \textcolor{black}{$I(Y(t_n);X_{1}(t_n),X_{2}(t_n))$}; the latter was computed following the formulation sketched in Subsect. "Linear parametric formulation" with \textcolor{black}{$T=Y(t_n)$ and $S_i=X_{i}(t_n)$}, after deriving the zero-lag covariance of the processes via solution of the Yule Walker equations of the VAR process (the procedure is illustrated in detail in \cite{mijatovic2024network}).


The results suggest that the rate of dynamic information shared by multivariate processes is deeply affected by the balance between instantaneous and time-lagged interactions.
When the three processes interact only at lag zero and do not exhibit self-dependencies ($c=0$), the spectral profiles of the MIR and redundancy rate measures are flat (Fig. \ref{fig:thsim_PID_vs_PIRD_spectral_functions}b-e, black lines) and the time-domain PIRD and zero-lag PID measures coincide (Fig. \ref{fig:thsim_PID_vs_PIRD_spectral_functions}f with $c=0$). Increasing the parameter $c$ determines modifications of the spectral profiles of the MIR functions (Fig. \ref{fig:thsim_PID_vs_PIRD_spectral_functions}b-d), due to the emergence of spectral peaks around $0.1$ and $0.3$ Hz induced by the self-dependencies rising in $X_1$ and $X_2$, as well as of causal interactions along the directions $X_1 \rightarrow Y$ and $X_2 \rightarrow Y$. As a result, the profile of the spectral redundancy rate is also modulated by $c$ (Fig. \ref{fig:thsim_PID_vs_PIRD_spectral_functions}e), as are the time-domain MIR and PIRD terms (Fig. \ref{fig:thsim_PID_vs_PIRD_spectral_functions}f, dashed lines).
The modification of the VAR parameters affects also the zero-lag PID (Fig. \ref{fig:thsim_PID_vs_PIRD_spectral_functions}f, solid lines), whose information atoms were modified in a substantially different way than those of the PIRD.
In fact, both redundant and synergistic contributions computed via PID ($\mathcal{R}$ and $\mathcal{S}$, solid lines in Fig. \ref{fig:thsim_PID_vs_PIRD_spectral_functions}f)  decrease towards zero at increasing $c$, while we rather expect an increase of synergy due to the emerging common child structure of the simulated system (Fig. \ref{fig:thsim_PID_vs_PIRD_spectral_functions}a). This effect is well evidenced by the PIRD, whose synergistic contribution increases with $c$ becoming far higher than the redundant contribution ($\mathcal{R}$ and $\mathcal{S}$, dashed lines in Fig. \ref{fig:thsim_PID_vs_PIRD_spectral_functions}f).
Moreover, the PIRD detects a rise of both the unique information rates relevant to the two sources ($\mathcal{U}_1$ and $\mathcal{U}_2$, dashed lines in Fig. \ref{fig:thsim_PID_vs_PIRD_spectral_functions}f), which is expected due to the emergence of causal interactions along the directions $X_1 \rightarrow Y$ and $X_2 \rightarrow Y$ with strength modulated by $c$.
However, the same is not true if the unique information is measured by the zero-lag PID exploiting the time domain definition of redundancy (\ref{redundancy_MMI}) based on the MMI-PID, confirming a known limitation of such PID which always forces to zero the unique information of the source sharing the lowest information with the target (in this case $\mathcal{U}_2$, dashed line in Fig. \ref{fig:thsim_PID_vs_PIRD_spectral_functions}f).

\subsection{\label{subsect:simulations_PID_transfer_entropies}Effects of changes in the network topology}

\begin{figure}
    \centering
\includegraphics[width=\columnwidth]{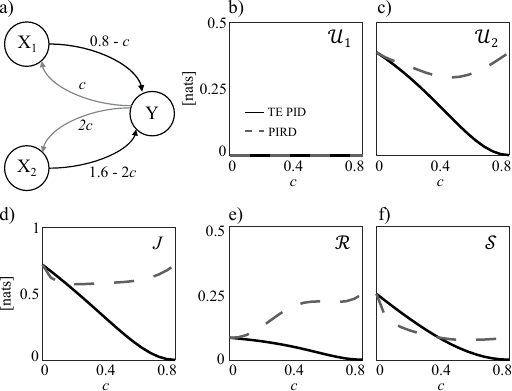}
    \caption{\textbf{Changes in the network topology have a profound impact on the multivariate information shared by multiple random processes.}    
    \textbf{a)} Simulation design, where $Y$ is the target process and $\{X_1,X_2\}$ is the group of sources; time-lagged interactions (solid black arrows) are set varying the parameter $c \in [0-0.8]$ to simulate a transition from purely unidirectional interactions directed from  $X_1, X_2$ to $Y$ when $c=0$ to purely unidirectional interactions directed from $Y$ to $X_1, X_2$ when $c=0.8$.
    (\textbf{b-e}) Profiles of the time domain MIR shared between the $Y$ and $\{X_1,X_2\}$ and the joint TE from $\{X_1,X_2\}$ to $Y$ (measuring overall interactions, here indicated as $J$) and of the unique ($\mathcal{U}_1$, $\mathcal{U}_2$), redundant ($\mathcal{R}$) and synergistic ($\mathcal{S}$) components of their decomposition measured using the PID applied to the TEs (continuous black lines) and the PIRD (dashed grey lines) as a function of the parameter $c$.}
    \label{fig:th_sim_PID_TE_vs_PIRD}
\end{figure}

Here, we investigate the time-domain behavior of the PIRD measures obtained decomposing the MIR between a target process $Y$ and two source processes $X_1,X_2$, compared with the same measures obtained from a PID applied to the joint TE from the two sources to the target, $T_{X_1,X_2 \rightarrow Y}$.
To this aim, we use a three-variate VAR process simulated with $f_s=1$ where the parameters related to lagged interactions from the sources to the target and vice versa are varied, in order to obtain different configurations of causal interactions.
Specifically, the 3-VAR process is defined as:
\begin{equation}
\begin{aligned}
&\textcolor{black}{Y(t_n)} = (0.8-c) \textcolor{black}{X_{1}(t_{n-1})} + (1.6-2c) \textcolor{black}{X_{2}(t_{n-1})} + \textcolor{black}{U_{Y}(t_n)}\\
&\textcolor{black}{X_{1}(t_n)} = c \textcolor{black}{Y(t_{n-1})} + \textcolor{black}{U_{X_1}(t_n)} \\
&\textcolor{black}{X_{2}(t_n)} = 2c \textcolor{black}{Y(t_{n-1})} + \textcolor{black}{U_{X_2}(t_n)}
\label{th_sim_TE_3VAR_equations}
\end{aligned}
\end{equation}
where $U_{Y}$, $U_{X_1}$ and $U_{X_2}$ are uncorrelated Gaussian white noises with zero mean and unit variance; the uncorrelation between the inputs ($\mathbf{\Sigma}_{\textcolor{black}{\textbf{U}}}=\mathbf{I}$) denotes absence of instantaneous correlations.
On the other hand, causal interactions are set at lag $k=1$, from $Y$ to both $X_1$ and $X_2$ with strength modulated directly by the parameter $c$, and from $X_1$ and $X_2$ to $Y$ with strength modulated inversely by the same parameter $c$. This setting allows for a progressive strengthening of the causal interactions directed from the target to the sources, and a progressive weakening of the causal interactions directed from the sources to the target, as $c$ increases in the range $[0-0.8]$. The simulation design is shown in Fig. \ref{fig:th_sim_PID_TE_vs_PIRD}a.

The PIRD was performed as in the first simulation, using the approach presented in Sect. "Formulation for Gaussian processes". Remarkably, since in this simulation of processes without self-dependencies the spectral MIR functions are flat, the time- and frequency-domain PIRD are equivalent and reduce to applying the MMI criterion to the MIR.
As regards the PID applied to the joint TE \textcolor{black}{$I(Y(t_n);X_{1}(t_{<n}),X_{2}(t_{<n})|Y(t_{<n}))$}, it was computed exploiting the formalism linking information-theoretic measures with Granger causality (GC) measures derived from linear parametric regression models developed for joint Gaussian processes \cite{barnett2009granger,barrett2010, faes2016information}. Specifically, the MMI criterion \cite{barrett2015exploration} was used to derive redundancy as the minimum information transferred from each individual source to the target; the latter was measured as half the value of the GC estimated from the VAR processes \cite{barnett2009granger}, and all GC terms were computed from the VAR parameters using sub-models \cite{faes2016information}. The resulting redundant TE ($\mathcal{R}$) was then used to derive the unique ($\mathcal{U}_1$, $\mathcal{U}_2$) and synergistic ($\mathcal{S}$) amounts of information transferred from $X$ to $Y$ according to PID rules.

The time domain behaviors of the PID terms are shown in Fig. \ref{fig:th_sim_PID_TE_vs_PIRD}b-f as a function of the parameters $c_1, c_2$ (continuous black lines).
The results evidence how the transition from the condition in which the target acts exclusively as a sink of information ($c=0$) to that in which it acts exclusively as a source of information ($c=0.8$) is thoroughly reflected by the PIRD but cannot be fully captured by the PID applied to the joint TE. Indeed, the latter cannot take time-lagged interactions directed from the target to the sources into account, and thus in the simulation where $c$ is increased it reflects only the decrease of the joint information transferred along the direction $\textcolor{black}{\textbf{X}} \rightarrow Y$ ($J$, Fig. \ref{fig:th_sim_PID_TE_vs_PIRD}d, solid line) and the consequent drop to zero of all PID terms ($\mathcal{U}_1$, $\mathcal{U}_2$, $\mathcal{R}$, $\mathcal{S}$, Fig. \ref{fig:th_sim_PID_TE_vs_PIRD}b,c,e,f, solid lines).
On the other hand, the PIRD is applied to the MIR shared between the sources and the target, a measure which is not directional as the TE and is thus sensible to the overall information flowing among the processes of the considered dynamic system. This is reflected by values of the MIR between $Y$ and $X_1,X_2$ which remain high as $c$ increases ($J$, Fig. \ref{fig:th_sim_PID_TE_vs_PIRD}d, dashed line), as also happens for the unique information rate of $X_2$ ($\mathcal{U}_2$, Fig. \ref{fig:th_sim_PID_TE_vs_PIRD}c, dashed line), and by values of redundancy rate and synergy rate shifting from the prevalence of synergy when $c=0$ to the prevalence of redundancy when $c=0.8$. The latter behavior reflects the modifications of the topological structure induced in the dynamic network, with links undergoing a transition from a common child configuration to a common drive configuration. 

\subsection{\label{subsect:simulations_3sources} Frequency-specific coarse-grained PIRD}

\begin{figure*}
    \centering
    \includegraphics{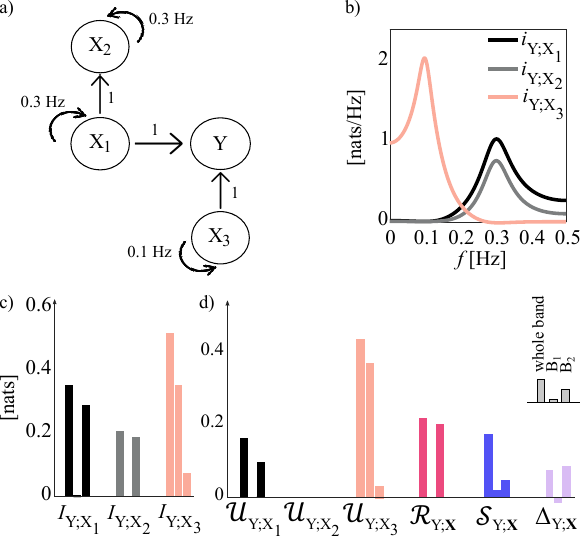}
    \caption{\textbf{Coexistence of redundant and synergistic characters of interactions in different spectral bands elicited by frequency-specific PIRD.}
    \textbf{(a)} Network structure, with $Y$ receiving from $X_1$, oscillating at 0.3 Hz, and $X_3$, oscillating at 0.1 Hz, and $X_1$ sending to $X_2$, oscillating at 0.3 Hz; all coupling coefficients are set to 1.
    \textbf{(b)} Spectral profiles of the pairwise MIR measures computed between the target $Y$ and each source $X_i$, $i=1,2,3$. 
    \textbf{(c,d)} Time-domain values of the pairwise MIR and of the unique ($\mathcal{U}_{Y;X_i}$, $i=1,2,3$), redundant ($\mathcal{R}_{Y;\textcolor{black}{\textbf{X}}}$) and synergistic ($\mathcal{S}_{Y;\textcolor{black}{\textbf{X}}}$) PIRD components integrated along the whole frequency axis ($\left[ 0 - f_s/2\right]$) (left bars), and withing the bands $B_1=\left[ 0.04 - 0.15\right]$ Hz (middle bars), $B_2=\left[ 0.15 - 0.4\right]$ Hz (right bars).
    The balance $\Delta_{Y;\textcolor{black}{\textbf{X}}}=\mathcal{R}_{Y;\textcolor{black}{\textbf{X}}}-\mathcal{S}_{Y;\textcolor{black}{\textbf{X}}}$ is also reported, indicating overall prevalence of redundancy and coexistence of redundancy and synergy in the two different bands $B_1$ and $B_2$.
    }
    \label{fig:thsim_mix_new}
\end{figure*}

In the last simulation, we provide an illustrative example of the full-frequency and band-limited PIRD applied to a network of four nodes where different high-order behaviors emerge at different frequencies.
We consider a VAR process simulated with $f_s=1$, where the lagged interactions between the target $Y$ and the three sources $\textcolor{black}{\textbf{X}}=\left[X_1,X_2,X_3\right]$ are set to induce a network topology with both common drive and common child structures (Fig. \ref{fig:thsim_mix_new}a). The VAR process is defined as:
\begin{equation}
\begin{aligned}
&\textcolor{black}{Y(t_n)} = \textcolor{black}{X_{1}(t_{n-1})} + \textcolor{black}{X_{3}(t_{n-1})} + \textcolor{black}{U_{Y}(t_n)} \\
&\textcolor{black}{X_{1}(t_n)} = \sum_{k=1}^{2} a_{1,k} \textcolor{black}{X_{1}(t_{n-k})} + \textcolor{black}{U_{X_1}(t_n)}  \\
&\textcolor{black}{X_{2}(t_n)} = \sum_{k=1}^{2} a_{2,k} \textcolor{black}{X_{2}(t_{n-k})} + \textcolor{black}{X_{1}(t_{n-1})} + \textcolor{black}{U_{X_2}(t_n)} \\
&\textcolor{black}{X_{3}(t_n)} = \sum_{k=1}^{2} a_{3,k} \textcolor{black}{X_{3}(t_{n-k})} + \textcolor{black}{U_{X_3}(t_n)} 
\label{th_sim_4VAR_equations}
\end{aligned}
\end{equation}
where $\textcolor{black}{\textbf{U}} =\left[ U_{Y},U_{X_1},U_{X_2},U_{X_3} \right]$ is a vector of four zero-mean independent Gaussian white noises with unit variance. The autonomous oscillations in the three source processes $X_1$, $X_2$, $X_3$ are obtained placing a pair of complex-conjugate poles, with modulus $\rho$ and phase $2\pi{f}$, in the complex plane representation of each process. 
Here, we set $\rho=0.8, f=0.3$ Hz for the processes $X_1$ and $X_2$, so that their autonomous dynamics are determined by the coefficients $a_{1,1}=a_{2,1}=-0.494; a_{1,2}=a_{2,2}=-0.64$; similarly, we set $\rho=0.9, f=0.1$ Hz for $X_3$, so that its autonomous dynamics are determined by the coefficients $a_{3,1}=1.456; a_{3,2}=-0.81$. Moreover, causal interactions are set from $X_1$ and $X_3$ to $Y$ and from $X_1$ to $X_2$ at lag $k=1$, with unitary strength. 

The PIRD was applied computing the spectral MIRs between each individual source and the target ($i_{Y;X_i}(f)$, $i=1,2,3$; Fig. \ref{fig:thsim_mix_new}b), as well as between groups of sources and the target (e.g., $i_{Y;X_1,X_2}(f)$); the spectral redundancy function was then computed as in (\ref{spectral_redundancy_rate_function_MMI}), and was exploited to retrieve the unique ($\mathcal{U}_{Y;X_1}$, $\mathcal{U}_{Y;X_2}$, $\mathcal{U}_{Y;X_3}$), redundant ($\mathcal{R}_{Y;\textcolor{black}{\textbf{X}}}$) and synergistic ($\mathcal{S}_{Y;\textcolor{black}{\textbf{X}}}$) information rates in the time domain; the latter were obtained as the whole-band integral of the correspondent spectral functions, as well as the integral taken along  two spectral bands centered around the simulated stochastic oscillations (i.e., $B_1 = \left[ 0.04 - 0.15 \right]$ Hz and $B_2 = \left[ 0.15 - 0.4 \right]$ Hz). 

The resulting time-domain values of the individual MIR terms and of the PIRD components are shown in Fig. \ref{fig:thsim_mix_new}c,d. The comparison highlights how, contrary to the pairwise MIR, the coarse-grained PIRD allows to disentangle the underlying network structure.
Indeed, whilst non-zero MIR values are detected between $Y$ and $X_2$ (Fig. \ref{fig:thsim_mix_new}c), the unique contribution of $X_2$ to $Y$ was null, and non-zero unique contributions  are correctly identified only in the presence of direct links (i.e., $\mathcal{U}_{Y;X_1}$, $\mathcal{U}_{Y;X_3}$). 
Remarkably, such non-zero unique contributions are mainly visible when assessed within the frequency bands for which oscillatory components are imposed (i.e., $B_2$ for $\mathcal{U}_{Y;X_1}$ and $B_1$ for $\mathcal{U}_{Y;X_3}$), thus confirming the important role played by the spectral representation of PIRD in the analysis of rhythmic processes.

The PIRD also favors quantification of redundancy and synergy related to the full dynamical structure of the analyzed processes, or to oscillations confined within the bands $B_1$ and $B_2$ (Fig. \ref{fig:thsim_mix_new}d). Redundancy arises typically from common drive (sub)structures where multiple copies of the same information are distributed, providing robustness \cite{luppi2024information}; e.g., here $X_1$ sends redundant information at $\sim 0.3$ Hz to both $X_2$ and $Y$, which is correctly detected by the significant values of $\mathcal{R}_{Y;\textcolor{black}{\textbf{X}}}$ within $B_2$.
On the other hand, synergistic informational circuits generally emerge from common child configurations, requiring a high degree of coordination between multiple parts of the system \cite{varley2023multivariate}; here, since $X_1$ and $X_3$ send information to $Y$ at different frequencies, synergy is detected in both bands $B_1$ and $B_2$.
The balance between synergy and redundancy assessed across the full spectrum, $\Delta_{Y;\textcolor{black}{\textbf{X}}}=\mathcal{R}_{Y;\textcolor{black}{\textbf{X}}} - \mathcal{S}_{Y;\textcolor{black}{\textbf{X}}}$, indicates an an overall prevalence of redundancy within the network. Interestingly, $\Delta_{Y;\textcolor{black}{\textbf{X}}}$ indicates the presence of net synergy when assessed within $B_1$, due to the fact that only the source $X_3$ transfers information to the target $Y$ in this band, and of net redundancy when assessed within $B_2$, due to the common drive role of the source $X_1$ in this band. 
Therefore, the combination of common drive and common child substructures, with the source processes oscillating at different frequencies, leads to the coexistence of synergistic and redundant modes of interplay in distinct spectral bands. These complex behaviors emerging at different time scales can be detected only using frequency-specific measures of redundancy and synergy, and can be better characterized separating redundant and synergistic contributions as guaranteed by the PIRD.

\section{\label{sect:applications}Application to physiological networks}

This section presents the application of the proposed framework to the network of physiological processes involved in the homeostatic control of arterial pressure and cerebral blood flow. The applicative context is very popular in the field of computational physiology, where the spontaneous variability of heart rate, arterial pressure, respiration and cerebral blood flow is widely studied to assess non-invasively the mechanisms involved in cardiovascular and cerebrovascular regulation, using a variety of time series analysis techniques including information decomposition methods \cite{zhang2000spontaneous, malpas2002neural, faes2013, faes2016information, porta2017quantifying, gelpi2022dynamic, sparacino2023method, mijatovic2024assessing, sparacino2024measuring}.

\subsection{\label{subsect:applications_data_acquisition}Experimental protocol} 

The application involves physiological time series reflecting th spontaneous variability of cerebral blood flow (CBFV), arterial pressure (AP), heart period (HP) and respiration (RESP). The analyzed time series belong to a database previously collected to study the short-term cardiovascular and cerebrovascular control responses to orthostatic challenge in subjects prone to neurally-mediated syncope 
via the analysis of spontaneous variability of systemic variables \cite{faes2013, bari2016}. The study included 13 subjects (age: $28\pm{9}$ years; 5 males) with previous history of unexplained syncope (reporting $>$ 3 syncope events in the previous 2 years)
, enrolled at the Neurology Division of Sacro Cuore Hospital, Negrar, Italy. The protocol consisted of 10 minutes of recording in the resting supine position, followed by prolonged stay in the 60$^\circ$ position reached after passive head-up tilt. All subjects experienced presyncope signs (i.e., a vasovagal episode characterized by hypotension and reflex bradycardia leading to partial loss of consciousness) during the tilt session; when signs were reported, the subject was returned to the resting position and a spontaneous recovery occurred. 

The acquired signals were the electrocardiogram (ECG, lead II), the continuous AP measured at the level of middle finger through a photopletysmographic device (Finapres, Enschede, The Netherlands), the CBFV signal measured at the level of the middle cerebral artery by means of a transcranial doppler ultrasonographic device (Multi-Dop T, Compumedics, San Juan Capistrano, CA, USA), and the respiratory amplitude signal measured through a thoracic impedance belt. From these signals, physiological time series were synchronously extracted on a beat-to beat basis starting from the heart period (HP), which was measured as the temporal distance between two consecutive R peaks of the ECG detected through a template matching algorithm \cite{bari2016}. 
Then, the $n^{\mathrm{th}}$ systolic AP ($SAP(n)$) was measured as the maximum of the AP signal inside the $n^{\mathrm{th}}$ HP ($HP(n)$). The $n^{\mathrm{th}}$ diastolic AP value ($DAP(n)$) was taken as the minimum AP between the occurrences of $SAP(n)$ and $SAP(n+1)$. The mean AP (MAP) values were computed by integrating the AP signal between the occurrences of $DAP(n-1)$ and $DAP(n)$ and, then, by dividing the result by the duration of the $n^{\mathrm{th}}$ diastolic interval (i.e., the time distance between the occurrences of $DAP(n-1)$ and $DAP(n)$). The mean CBFV (MCBFV) values were computed by integrating the CBFV signal between the diastolic values (i.e., the minima of the CBFV close to the occurrences of $DAP(n-1)$ and $DAP(n)$) and, then, by dividing the result by the time distance between the two diastolic values. Finally the $n^{\mathrm{th}}$ RESP value (i.e., $RESP(n)$) was computed sampling the respiration signal on the $n^{\mathrm{th}}$ R peak of the ECG.

The beat-to-beat variability series of HP, MAP, MCBFV and RESP, herein referred respectively as $H$, $M$, $F$ and $R$, were then produced as the sequences of consecutive values collected during three stationary time windows of length $L=250$ beats during the following physiological conditions \cite{faes2013, bari2016}: (i) supine rest (RS); (ii) early tilt (ET), starting after the onset of the head-up tilt maneuver, excluding transient changes of the physiological variables; and (iii) late tilt (LT), 
occurring just before the pressure decrease due to presyncope (start at $16 \pm 8$ min after the head-up tilt). 
Further information about the experimental protocol, signal acquisition and variability series extraction can be found in \cite{faes2013, bari2016}.

\subsection{\label{subsect:applications_data_analysis}Data and statistical analysis}
The four physiological time series described above were taken as a realization of the random processes \textit{F}, \textit{M}, \textit{H} and \textit{R}. These processes were investigated to decompose the rate of information shared dynamically between the target $Y=F$ and the set of source processes $\textcolor{black}{\textbf{X}}=\{ H,M,R \}$, with $H=X_1, M=X_2, R=X_3$, with the aim of investigating the mechanisms underlying the short-term dynamic regulation of the cerebral blood flow in resting and stress conditions.
Each series was first detrended with an AR high-pass filter with zero phase (cutoff frequency 0.015 cycles/beat) \cite{nollo2000}. Then, a VAR model in the form of (\ref{VAR_model}) was fitted to the four time series; model identification was performed via the ordinary least-squares approach \cite{lutkepohl2005new}, setting the model order $p$ according to the Akaike Information Criterion (AIC) for each subject (with maximum model order equal to 14) \cite{akaike1974}. After VAR identification, computation of time and frequency domain interaction measures of mutual, unique, redundant and synergistic information rates was performed from the estimated model parameters and spectra of the processes. Spectral analysis was performed assuming the series as uniformly sampled with the mean HP taken as the sampling period. Specifically, the spectral MIRs between the target and groups of sources were computed as in (\ref{spectralMIR}); then, according to coarse-graining PIRD, the unique, redundant and synergistic information rates calculated from the spectral redundancy rate in (\ref{spectral_redundancy_rate_function_MMI}) were computed through integration within the low frequency (LF, $\left[ 0.04 - 0.15 \right]$ Hz) and high frequency (HF, $\left[ 0.15 - 0.4 \right]$ Hz) bands of the spectrum to analyze specific rhythms with physiological meaning \cite{malpas2002neural}, as well as through integration over all frequencies, to get overall time-domain measures. 

Given the small size of the surveyed population, non-parametric statistical tests were applied to assess statistically significant differences between indexes evaluated in the three phases of the experimental protocol, i.e., RS, ET and LT conditions. Specifically, the Wilcoxon signed rank test for paired data was applied on the MIRs shared between the target and each source, as well as on the PIRD terms (i.e., the joint MIR shared between the target and all the sources, the unique, redundant and synergistic information rates), evaluated in the time domain and along the LF and HF bands of the spectrum during the three phases of the protocol. Further, the statistical differences among pairs of unique information rate measures, as well as between redundancy and synergy, were tested in the time domain, LF and HF bands for each experimental condition via Wilcoxon signed rank test for paired data with Bonferroni-Holm correction for multiple comparisons. For all the statistical tests, the significance level was set to $0.05$. 

\begin{figure*}
    \centering
    \includegraphics[width=\textwidth]
    {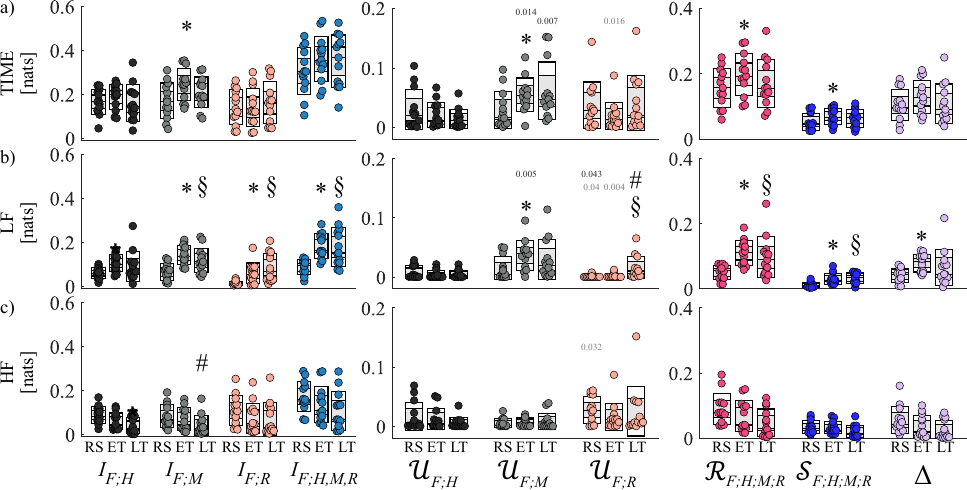}
    \caption{\textbf{Coarse-grained PIRD applied to the physiological network of mean cerebral blood flow velocity ($F$), mean arterial pressure ($M$), heart period ($H$) and respiration ($R$) assessed in patients prone to develop postural-related syncope.}
    The mutual (left column: $I_{F;H}$ (black dots), $I_{F;M}$ (grey dots), $I_{F;R}$ (pink dots), $I_{F;H,M,R}$ (light blue dots)), unique (middle column: $\mathcal{U}_{F;H}$ (black dots), $\mathcal{U}_{F;M}$ (grey dots), $\mathcal{U}_{F;R}$ (pink dots)), redundant (right column: $\mathcal{R}_{F;H;M;R}$, magenta dots) and synergistic (right column: $\mathcal{S}_{F;H;M;R}$, blue dots) information rates shared between the target $F$ and the sources $\{ H,M,R \}$ are computed in the RS (left boxplots), ET (middle boxplots) and LT (right boxplots) conditions along \textbf{a)} the whole frequency axis (TIME), \textbf{b)} the LF and \textbf{c)} the HF bands of the spectrum, taking the spectral redundancy rate function as in (\ref{spectral_redundancy_rate_function_MMI}). The redundancy-synergy balance $\Delta$ is shown as boxplot distributions and individual values (purple dots) in the right column of panels a) (TIME), b) (LF) and c) (HF). The statistical differences among pairs of unique information rates (middle panels) were tested via Wilcoxon signed rank test for paired data ($p<0.05$) with Bonferroni-Holm correction for multiple comparisons: significant p-values are shown above the boxplot distributions with colors indicating the distribution to be compared with (i.e., black if compared with $\mathcal{U}_{F;H}$, grey if compared with $\mathcal{U}_{F;M}$). Wilcoxon signed rank test for paired data, $p<0.05$: $\ast$, RS vs. ET; $\S$, RS vs. LT; $\#$, ET vs. LT.}
    \label{fig:appl_negrar}
\end{figure*}

\subsection{\label{subsect:applications_results}Results and discussion}
Results are shown in Fig. \ref{fig:appl_negrar} as boxplot distributions and individual values of the PIRD measures computed in the RS, ET and LT conditions. 
The MIRs $(I_{F;H},I_{F;M},I_{F;R},I_{F;H,M,R})$ and unique information rates $(\mathcal{U}_{F;H}, \mathcal{U}_{F;M}, \mathcal{U}_{F;R})$ assessed in the time domain (panels \textbf{a)}) and within the LF (panels \textbf{b)}) and HF (panels \textbf{c)}) bands of the spectrum are depicted in the left and middle columns, respectively, while the redundant and synergistic contributions and their balance ($\Delta$) are shown in the right column.

The transition to ET induced an increase of $I_{F;M}$ and $\mathcal{U}_{F;M}$ mainly visible in the LF band of the spectrum (Fig. \ref{fig:appl_negrar}a,b, left and middle columns), in line with previous observations indicating that the increase of the information transfer from \textit{M} to \textit{F} is related to altered cerebral autoregulation (CA) in syncope patients \cite{bari2017}. The LT-induced decrease of $I_{F;M}$ in HF is probably related to the modulating effect of respiration \cite{bari2016}, and thus is not detected by the unique contribution of arterial pressure ($\mathcal{U}_{F;M}$, panel c, middle column).
As regards the interactions between cerebral blood flow and respiration, we found that the MIR $I_{F;R}$ increases during the early phase of tilt in LF probably due to confounding effects of $H$, $M$, while RS and ET values of $\mathcal{U}_{F;R}$ are very close to zero for most of subjects (Fig. \ref{fig:appl_negrar}b, left and middle columns), suggesting that the unique information shared between $R$ and $F$ may be negligible during this phase of the protocol. This result is further confirmed by the significantly lower values of $\mathcal{U}_{F;R}$ than $\mathcal{U}_{F;H}$ and $\mathcal{U}_{F;M}$ in the RS and ET phases. However, a significant increase of the unique information rate shared between the respiratory and the cerebral system in the LF band is observed in LT, thus shedding light on the remarkable role of respiration in influencing the dynamics of CBFV suddenly before the occurrence of syncope \cite{porta2008influence, bari2016}.
It is worth noting that the tilt-induced significant increase of cerebro-vascular and cerebro-respiratory interactions in the LF band is well visible as significant increases of the joint MIR $I_{F;H,M,R}$ (Fig. \ref{fig:appl_negrar}b, left column), which however is not a \textit{source-specific} measure and thus cannot distinct between the pathways of information flow.

A major result is related to the interactions between cerebral and cardiac processes, which indeed do not change in response to the orthostatic challenge within the investigated network ($I_{F;H}$, $\mathcal{U}_{F;H}$, black distributions of left and middle panels), thus implying invariance of these relationships with the postural stress.
Nonetheless, besides being significantly lower than $\mathcal{U}_{F;M}$ during tilt as suggested by p-values $<0.05$ (Fig. \ref{fig:appl_negrar}a, middle column), a result that emerges better in the LF band (Fig. \ref{fig:appl_negrar}b, middle column), the role played by heart rate in influencing CBFV variability in this group of subjects cannot be marked as negligible throughout the experimental protocol outlined in the study. Instead, we remark here the prevalence of the cerebro-vascular interactions between cerebral blood flow and arterial pressure.
Overall, what emerges from the analysis of mutual and unique information rates is that the latter is able to capture direct mechanisms of interaction which can be masked by the presence of other unobserved variables in the MIR measure: looking at the unique contributions, only the $M-F$ interactions seem to be significantly affected by the early postural stress in the LF band, while respiration may play a role in shaping CBFV dynamics in the late phase of tilt.

In addition, the significant increase of the redundant and synergistic information rates assessed in the LF band moving from RS to ET and LT, with redundancy always significantly higher than synergy as documented by statistical tests, suggests that the orthostatic stress is responsible for the emergence of predominantly redundant patterns of interaction between cardiovascular and cardiorespiratory processes sharing information with the cerebral flow velocity taken as the target process. The significant increase of $\Delta$ in the early phase of tilt (Fig. \ref{fig:appl_negrar}b, right panel) remarks the importance of LF oscillations within the network and the LF-dependent nature of the interactions among the investigated signals.

Overall, these results confirm the redundant nature of cardiovascular and cerebrovascular interactions previously reported for similar triplets of physiological processes \cite{faes2016information, porta2017quantifying, faes2022new}, and document the relevance of separating LF and HF contributions to elicit the different roles of heart rate, arterial pressure and respiration on cardiovascular and cerebrovascular interactions. Moreover, the significant increase of redundancy with tilt suggests that these redundant effects are enhanced during postural stress, likely as a consequence of sympathetic activation and vagal withdrawal \cite{malpas2002neural,porta2017quantifying}.
Therefore, the evidence that cardiovascular and cerebrovascular interactions occur through the coupling of rhythms in different frequency bands with different physiological meaning make the proposed spectral PIRD eligible to probe high-order interactions in these networks \cite{faes2021information}. 

\section*{Conclusions}
In this work, we introduced the partial information rate decomposition (PIRD) framework by extending partial information decomposition (PID) to random processes with temporal statistical structure. We provided a decomposition of the information shared dynamically between a target process and a set of source processes into unique, redundant, and synergistic contributions over time, by utilizing the concept of rate of mutual information shared instead of the well-known mutual information. A key innovation is the introduction of a spectral redundancy rate function, formulated exploiting a pointwise definition of redundancy in the spectral domain, thus enabling a frequency-specific decomposition of information interactions. Furthermore, we introduced a coarse-grained representation of information decomposition, which aggregates partial information atoms into a principled structure that remains scalable and interpretable even in high-dimensional dynamic systems. By integrating spectral analysis and coarse-graining, our framework provides a refined and computationally efficient partition of information flow in complex networks.

Through benchmark simulations, we demonstrated how conventional PID methods can fail to properly decompose high-order interactions in dynamic systems, due to a missing or partial account of temporal correlations. In contrast, our framework provides a more accurate and interpretable decomposition of multivariate information flow in dynamic systems of random processes. Additionally, the application of the PIRD to a physiological network involving cerebrovascular and cardiovascular variables revealed that frequency-dependent redundant information exchange plays a crucial role in the response to postural stress, as well as that cardiovascular and cerebrovascular interactions occur through the coupling of rhythms in different frequency bands with different physiological meaning. This emphasizes the importance of availing of spectrally resolved information decomposition methods in the presence of networks of random processes with relevant oscillatory content, which can exhibit the coexistence of redundant and synergistic higher-order interactions within different frequency bands.

Overall, by integrating temporal structure and frequency-specific dependencies, PIRD offers a more refined and comprehensive approach for studying multivariate interactions in neuroscience, biology, engineering, and beyond. Future works will explore further refinements in redundancy definitions \textcolor{black}{to be used in PIRD schemes involving purely discrete \cite{williams2010nonnegative}, continuous \cite{ehrlich2024partial} and mixed \cite{bara2024partial} random processes and different corresponding examples of dynamic interactions where to test them \cite{timme2014synergy,barnett2015computational,siggiridou2019evaluation}, as well as investigate} their applicability to real-world networked systems with nonlinear dependencies.

\textcolor{black}{\section*{Appendix A}
This appendix provides extended mathematical derivations to support and expand upon the new definition of spectral MMI redundancy rate discussed in Sect. "Frequency-domain PIRD".
Derivations are given considering a dynamic network of $M+1$ random processes $\{Y,X_1,\ldots,X_M\}$ where $Y$ is the target $\textbf{X}=\{X_1,\ldots,X_M\}$ is the group of $M$ sources.}

\textcolor{black}{\subsection*{Appendix A.1}
In this section, we demonstrate that the SMMI redundancy rate defined in (\ref{time_redundancy_rate_SMMI}) is less or equal than the MMI redundancy rate given in (\ref{redundancy_MMI}), with equality holding in the absence of temporal correlations.}

\textcolor{black}{Let us consider the generic $\alpha^{\mathrm{th}}$ atom of the lattice structure, $\alpha=\{\alpha_1,\ldots,\alpha_J \}$, with cardinality $J$; the $j^{\mathrm{th}}$ element of the atom (i.e., $\alpha_j$) contains $K_j$ sources; e.g., the atom $\{\{1\}\{23\}\}$ has $J=2$ and $K_1=1, K_2=2$.}
\textcolor{black}{We remark that the spectral redundancy rate function associated to the $\alpha^{\mathrm{th}}$ atom is computed at each frequency $\omega$ as in (\ref{spectral_redundancy_rate_function_MMI}). Integration of the MIR spectral profiles $i_{Y;\textbf{X}_{\alpha_j}}(\omega)$, $j=1,\ldots,J$, in the range $\left[ -\pi, \pi\right]$ returns the time-domain MIRs $I_{Y;\textbf{X}_{\alpha_j}}$ according to (\ref{spectralMIR_integral}); the time-domain SMMI redundancy rate, $I^{\mathrm{SMMI}}_{Y;\textbf{X}_{\alpha}}$, is derived by (\ref{time_redundancy_rate_SMMI}). Moreover, taking the minimum of the set $\{I_{Y;\textbf{X}_{\alpha_j}}\}$, $j=1,\ldots,J$, yields the MMI redundancy rate $I^{\mathrm{MMI}}_{Y;\textbf{X}_{\alpha}}$ as in (\ref{redundancy_MMI}). In light of this, we demonstrate the following proposition.}\\

\textcolor{black}{\textbf{Proposition A.1.1.} $I^{\mathrm{SMMI}}_{Y;\textbf{X}_{\alpha}}\leq I^{\mathrm{MMI}}_{Y;\textbf{X}_{\alpha}}$.}\\

\textcolor{black}{\textbf{Proof.} Since the minimum spectral redundancy rate function is taken pointwise in the frequency domain, we have that 
\[ i^{\mathrm{SMMI}}_{Y;\textbf{X}_{\alpha}}(\omega) \leq i_{Y;\textbf{X}_{\alpha_j}}(\omega) \]
$\forall \omega \in \left[ -\pi, \pi\right]$,  $\forall j=1,\ldots, J$. If the functions $i_{Y;\textbf{X}_{\alpha_j}}(\omega)$ are non-negative and integrable in $\left[ -\pi, \pi \right]$ (e.g., for Gaussian processes \cite{nedungadi2011block, geweke1982measurement}), $i^{\mathrm{SMMI}}_{Y;\textbf{X}_{\alpha}}(\omega)$ is also non-negative and integrable in $\left[ -\pi, \pi \right]$ \cite{rudin1976pma}, resulting in
\[ \frac{1}{2 \pi} \int_{-\pi}^{\pi} i^{\mathrm{SMMI}}_{Y;\textbf{X}_{\alpha}}(\omega) d\omega \leq \frac{1}{2 \pi} \int_{-\pi}^{\pi} i_{Y;\textbf{X}_{\alpha_j}}(\omega) d\omega \]
$\forall j=1,\ldots, J$, due to the property of \textit{monotonicity of the integral} (i.e., the integral preserves order \cite{rudin1976pma}). This implies that
\[ \frac{1}{2 \pi} \int_{-\pi}^{\pi} i^{\mathrm{SMMI}}_{Y;\textbf{X}_{\alpha}}(\omega) d\omega \leq \min_{j=1,\ldots,J} \frac{1}{2 \pi} \int_{-\pi}^{\pi} i_{Y;\textbf{X}_{\alpha_j}}(\omega) d\omega, \]
i.e.,
\[ I^{\mathrm{SMMI}}_{Y;\textbf{X}_{\alpha}} \leq \min_{j=1,\ldots,J} I_{Y,\textbf{X}_{\alpha_j}}, \]
ultimately demonstrating the stated proposition.}

\textcolor{black}{Equality holds in the absence of temporal correlations, i.e., when the random processes $\{Y, X_1,\ldots,X_M \}$ reduce to i.i.d. random variables. If this is the case, the auto- and cross-correlation functions are Dirac delta functions, whose Fourier transforms return constant (cross-)spectral density matrices equal to the (co)variances of the processes, i.e.,
\[ P_Y(\omega) = \sigma^2_Y; \quad \textbf{P}_{\textbf{X}_{\alpha_j}}(\omega) = \mathbf{\Sigma}_{\textbf{X}_{\alpha_j}}; \quad \textbf{P}_{Y\textbf{X}_{\alpha_j}}(\omega) = \mathbf{\Sigma}_{Y;\textbf{X}_{\alpha_j}},\]
where $\sigma^2_Y$ is the variance of the target, $\mathbf{\Sigma}_{\textbf{X}_{\alpha_j}}$ is the $K_j \times K_j$ auto-covariance matrix of $\textbf{X}_{\alpha_j}$ and $\mathbf{\Sigma}_{Y;\textbf{X}_{\alpha_j}}$ is the $1 \times K_j$ cross-covariance of $Y$ and $\textbf{X}_{\alpha_j}$.
Under the hypothesis of gaussianity, the MIR between the variables $Y$ and $\textbf{X}_{\alpha_j}$ reduces to $I_{Y;\textbf{X}_{\alpha_j}} = - \frac{1}{2} \log \big( 1 - \rho^2(Y;\textbf{X}_{\alpha_j}) \big)$, where
\[
\rho^2(Y;\textbf{X}_{\alpha_j}) = \frac{ \mathbf{\Sigma}_{Y;\textbf{X}_{\alpha_j}} \mathbf{\Sigma}^{-1}_{\textbf{X}_{\alpha_j}} \mathbf{\Sigma}^{\intercal}_{Y;\textbf{X}_{\alpha_j}} }{\sigma^2_Y} = 1- \frac{|\mathbf{\Sigma}_{[Y\textbf{X}_{\alpha_j}]}|}{\sigma^2_Y |\mathbf{\Sigma}_{\textbf{X}_{\alpha_j}}|}
\]
is the squared correlation coefficient, with $\mathbf{\Sigma}_{[Y\textbf{X}_{\alpha_j}]} = \begin{pmatrix}
    \sigma^2_Y & \mathbf{\Sigma}_{Y;\textbf{X}_{\alpha_j}} \\
    \mathbf{\Sigma}^{\intercal}_{Y;\textbf{X}_{\alpha_j}} & \mathbf{\Sigma}_{\textbf{X}_{\alpha_j}}
\end{pmatrix}$
the full joint covariance matrix. This leads to
\[
I_{Y;\textbf{X}_{\alpha_j}} = \frac{1}{2} \log \biggr( \frac{\sigma^2_Y |\mathbf{\Sigma}_{\textbf{X}_{\alpha_j}}|}{|\mathbf{\Sigma}_{[Y\textbf{X}_{\alpha_j}]}|} \biggr),
\]
which yields
\[ I^{\mathrm{MMI}}_{Y;\textbf{X}_{\alpha}} = \min_{j=1,\ldots,J} I_{Y;\textbf{X}_{\alpha_j}} = \min_{j=1,\ldots,J} \frac{1}{2} \log \biggr( \frac{\sigma^2_Y |\mathbf{\Sigma}_{\textbf{X}_{\alpha_j}}|}{|\mathbf{\Sigma}_{[Y\textbf{X}_{\alpha_j}]}|} \biggr). \]
In line with (\ref{spectralMIR_alphaj}) and the statements above, the spectral MIR does not depend on the frequency $\omega$, displaying a flat profile $\forall \omega \in \left[ -\pi, \pi \right]$, and can be written as 
\[
i_{Y;\textbf{X}_{\alpha_j}}(\omega) = \frac{1}{2} \log \biggr( \frac{\sigma^2_Y |\mathbf{\Sigma}_{\textbf{X}_{\alpha_j}}|}{|\mathbf{\Sigma}_{[Y\textbf{X}_{\alpha_j}]}|} \biggr);
\]
it follows that the spectral redundancy rate function $i^{\mathrm{SMMI}}_{Y;\textbf{X}_{\alpha}}$ does not vary with $\omega$, yielding 
\[ I^{\mathrm{SMMI}}_{Y;\textbf{X}_{\alpha}} = \min_{j=1,\ldots,J} \frac{1}{2} \log \biggr( \frac{\sigma^2_Y |\mathbf{\Sigma}_{\textbf{X}_{\alpha_j}}|}{|\mathbf{\Sigma}_{[Y\textbf{X}_{\alpha_j}]}|} \biggr),\]
ultimately demonstrating that $I^{\mathrm{MMI}}_{Y;\textbf{X}_{\alpha}} = I^{\mathrm{SMMI}}_{Y;\textbf{X}_{\alpha}}$.}

\textcolor{black}{\subsection*{Appendix A.2}
In this section, we demonstrate that moving from the PIRD defined on the pointwise level (frequency-specific) to that defined on the process-level (time domain) is straightforward exploiting spectral integration and basic properties of the integral.}\\

\textcolor{black}{\textbf{Proposition A.2.1.} 
Solving the PIRD as 
\[ \textbf{(a)} \quad I^{\delta}_{Y;\textbf{X}_{\alpha}} = \frac{1}{2 \pi} \int_{-\pi}^{\pi} i^{\delta}_{Y;\textbf{X}_{\alpha}}(\omega) d\omega, \]
where, according to (\ref{atom_information_rate_function_PIRD}),
\[ i^{\delta}_{Y;\textbf{X}_{\alpha}}(\omega) = i^{\cap}_{Y;\textbf{X}_{\alpha}}(\omega) - \sum_{\beta \prec \alpha} i^{\delta}_{Y;\textbf{X}_{\beta}}(\omega), \]
is equivalent to solve the PIRD as
\[ \textbf{(b)} \quad I^{\delta}_{Y;\textbf{X}_{\alpha}} = I^{\cap}_{Y;\textbf{X}_{\alpha}} - \sum_{\beta \prec \alpha} I^{\delta}_{Y;\textbf{X}_{\beta}}, \]
where, according to (\ref{time_redundancy_rate_SMMI}),
\[ I^{\cap}_{Y;\textbf{X}_{\alpha}} = \frac{1}{2 \pi} \int_{-\pi}^{\pi} i^{\mathrm{SMMI}}_{Y;\textbf{X}_{\alpha}}(\omega) d\omega.\]}\\

\textcolor{black}{\textbf{Proof.} The equivalence between statements \textbf{(a)} and \textbf{(b)} follows directly by applying the \textit{linearity of the integral} \cite{rudin1976pma}: \[ \int \big( f(x) - \sum_i g_i(x) \big) dx = \int f(x) dx - \sum_i \int g_i(x) dx, \] where $f(x)=i^{\cap}_{Y;\textbf{X}_{\alpha}}(\omega)$ and $g(x)=i^{\delta}_{Y;\textbf{X}_{\alpha}}(\omega)$. In this case, we have:
\[ I^{\delta}_{Y;\textbf{X}_{\alpha}} = \frac{1}{2\pi} \int_{-\pi}^{\pi} \big[ i^{\cap}_{Y;\textbf{X}_{\alpha}}(\omega) - \sum_{\beta \prec \alpha} i^{\delta}_{Y;\textbf{X}_{\beta}}(\omega) \big] d\omega
\]
\[
= \frac{1}{2\pi} \int_{-\pi}^{\pi} i^{\cap}_{Y;\textbf{X}_{\alpha}}(\omega) d\omega - \sum_{\beta \prec \alpha} \left( \frac{1}{2\pi} \int_{-\pi}^{\pi} i^{\delta}_{Y;\textbf{X}_{\beta}}(\omega) d\omega \right)
\]
\[
= I^{\cap}_{Y;\textbf{X}_{\alpha}} - \sum_{\beta \prec \alpha} I^{\delta}_{Y;\textbf{X}_{\beta}},
\]
which confirms that the frequency-domain decomposition \textbf{(a)} and the time-domain decomposition \textbf{(b)} are equivalent.}

\textcolor{black}{\subsection*{Appendix A.3}
In this section, we demonstrate that the SMMI spectral redundancy rate defined by (\ref{spectral_redundancy_rate_function_MMI}, \ref{spectralMIR_alphaj}) satisfies the properties of weak symmetry (Proposition A.3.1.), self-redundancy (Proposition A.3.2.), monotonicity (Proposition A.3.3.) and non-negativity (Proposition A.3.4.) proposed by Williams and Beer \cite{williams2010nonnegative}.}\\

\textcolor{black}{\textbf{Proposition A.3.1.} (\textit{weak symmetry}) The SMMI spectral redundancy rate 
is invariant with the ordering of the sources, i.e., $i^{\mathrm{SMMI}}_{Y;\textbf{X}_{\{\alpha_1,\ldots,\alpha_J\}}}(\omega)$ is symmetric w.r.t. the $\alpha_j, j=1,\ldots,J$.}\\

\textcolor{black}{\textbf{Proof.} The proposition follows immediately from the property of commutativity of the minimum operator applied to (\ref{spectral_redundancy_rate_function_MMI}).}\\

\textcolor{black}{\textbf{Proposition A.3.2.} (\textit{self-redundancy}) The SMMI spectral redundancy rate applied to a single collection of sources reduces to the spectral MIR between the sources and the target, i.e. $i^{\mathrm{SMMI}}_{Y;\textbf{X}_{\alpha}}(\omega)=i_{Y;\textbf{X}_{\alpha}}(\omega)$ when $J=|\alpha|=1$.}\\


\textcolor{black}{\textbf{Proof.} The proposition follows immediately from (\ref{spectral_redundancy_rate_function_MMI}) applied with $J=|\alpha|=1$.}\\


\textcolor{black}{\textbf{Proposition A.3.3.} (\textit{monotonicity} and \textit{subset equality}) The SMMI spectral redundancy rate does not increase when adding sources, i.e., $i^{\mathrm{SMMI}}_{Y;\textbf{X}_{\{\alpha_1,\ldots,\alpha_{J-1},\alpha_J\}}}(\omega) \leq i^{\mathrm{SMMI}}_{Y;\textbf{X}_{\{\alpha_1,\ldots,\alpha_{J-1}\}}}(\omega)$, with equality if $\alpha_{J-1} \subseteq \alpha_J$.}\\


\textcolor{black}{\textbf{Proof.} Let us denote:
\begin{itemize}
    \item $\alpha' = \{\alpha_1, \ldots, \alpha_{J-1}\}$,
    \item $\alpha'' = \alpha \cup \{\alpha_J\} = \{\alpha_1, \ldots, \alpha_{J-1}, \alpha_J\}$.
\end{itemize}
The spectral redundancy rate function over the expanded collection is given by:
\begin{multline*}
i^{\cap}_{Y; \textbf{X}_{\alpha''}}(\omega) = \min \{ i_{Y; \textbf{X}_{\alpha_1}}(\omega), \ldots, \\
i_{Y; \textbf{X}_{\alpha_{J-1}}}(\omega), i_{Y; \textbf{X}_{\alpha_J}}(\omega) \}.
\end{multline*}
Similarly, the spectral redundancy rate function over the original collection is:
\[
i^{\cap}_{Y; \textbf{X}_{\alpha'}}(\omega) = \min \{ i_{Y; \textbf{X}_{\alpha_1}}(\omega), \ldots, i_{Y; \textbf{X}_{\alpha_{J-1}}}(\omega) \}.
\]
Since taking the minimum over a larger set of values can only decrease or preserve the original minimum value, it follows that
\[
i^{\cap}_{Y; \textbf{X}_{\alpha''}}(\omega) \leq i^{\cap}_{Y; \textbf{X}_{\alpha'}}(\omega),
\]
thus proving monotonicity.}\\
\textcolor{black}{Subset equality, $i^{\cap}_{Y; \textbf{X}_{\alpha''}}(\omega) = i^{\cap}_{Y; \textbf{X}_{\alpha'}}(\omega)$ if $\alpha_{J-1} \subseteq \alpha_J$, holds thanks to the use of the minimum operator if 
\[
i_{Y;\textbf{X}_{\alpha_J}}(\omega) \geq i_{Y;\textbf{X}_{\alpha_{J-1}}}(\omega);
\]
the latter follows from the application of the chain rule to the spectral MIR $i_{Y;\textbf{X}_{\alpha_J}}(\omega)$. Specifically, denoting $\alpha_{J-1}=\{\alpha_{J-1}^1,\ldots,\alpha_{J-1}^{K_{J-1}} \}$ and $\alpha_{J}=\{\alpha_{J-1},\alpha_J^{K_{{J-1}}+1},\ldots,\alpha_J^{K_J} \}$, the chain rule reads
\[i_{Y;\textbf{X}_{\alpha_J}}(\omega)=i_{Y;\textbf{X}_{\alpha_{J-1}}}+ i_{Y;\textbf{X}_{\alpha_J^{{K_{{J-1}}+1}}},\ldots,\textbf{X}_{\alpha_J^{{K_J}}}|\textbf{X}_{\alpha_{J-1}}}(\omega),
\]
which proves that $i_{Y;\textbf{X}_{\alpha_J}}(\omega) \geq i_{Y;\textbf{X}_{\alpha_{J-1}}}(\omega)$ given the non-negativity of $i_{Y;\textbf{X}_{\alpha_J^{{K_{{J-1}}+1}}},\ldots,\textbf{X}_{\alpha_J^{{K_J}}}|\textbf{X}_{\alpha_{J-1}}}(\omega)$.
}\\

\textcolor{black}{\textbf{Proposition A.3.4.} (\textit{non-negativity}) The SMMI spectral redundancy rate is non negative, i.e. $i^{\mathrm{SMMI}}_{Y;\textbf{X}_{\alpha}}(\omega) \geq 0$ $\forall \omega \in [-\pi, \pi]$.
}\\

\textcolor{black}{\textbf{Proof.} The proof is given showing that each spectral MIR defined in (\ref{spectralMIR_alphaj}) at varying $j=1,\ldots,J$ is non-negative.
Applying the generalized Hadamard's inequality \cite{friedland1975generalised} to the spectral density matrix (\ref{PSDmatrixYXalphaj}) yields 
\[
|\textbf{P}_{[Y\textbf{X}_{\alpha_j}]}(\omega)|\leq P_{Y}(\omega)|\textbf{P}_{\textbf{X}_{\alpha_j}}(\omega)|,
\]
which corresponds to
\[i_{Y;\textcolor{black}{\textbf{X}}_{\alpha_j}}(\omega)=\frac{1}{2}\log \frac{|\textbf{P}_{\textcolor{black}{\textbf{X}}_{\alpha_j}}(\omega)|P_{Y}(\omega) }{|\textbf{P}_{[Y\textbf{X}_{\alpha_j}]}(\omega)|}\geq 0,
\]
thus completing the proof.
}



\section*{Acknowledgments}
This work was supported by the project “HONEST - High-Order Dynamical Networks in Computational Neuroscience and Physiology: an Information-Theoretic Framework”, Italian Ministry of University and Research (funded by MUR, PRIN 2022, code 2022YMHNPY, CUP: B53D23003020006).


\bibliography{biblio_Paper}


\end{document}